\documentclass[journal]{IEEEtran}
\IEEEoverridecommandlockouts
\usepackage{amsmath}
\usepackage{amsfonts}
\usepackage{amssymb}
\usepackage{xcolor}

\usepackage[english]{babel}
\usepackage{amsthm}
\newtheorem{remark}{Remark}

\newtheorem{lemma}{Lemma}

\newtheorem{proposition}{Proposition}
\usepackage{esint}
\usepackage{algorithm}
\usepackage{algorithmic}
\usepackage{array}
\usepackage[caption=false,font=normalsize,labelfont=sf,textfont=sf]{subfig}
\usepackage{textcomp}
\usepackage{stfloats}
\usepackage{multirow}
\usepackage{booktabs}
\usepackage{etoolbox}
\usepackage{geometry}
\usepackage{pstricks}

\usepackage{svg}

\usepackage{url}
\usepackage{verbatim}

\usepackage{cite}
\def\BibTeX{{\rm B\kern-.05em{\sc i\kern-.025em b}\kern-.08em
    T\kern-.1667em\lower.7ex\hbox{E}\kern-.125emX}}
\usepackage{subfloat}
\usepackage{tikz}
\usetikzlibrary{graphs, arrows, positioning, quotes, shapes.geometric, arrows.meta}
\usetikzlibrary{arrows.meta, positioning, calc, backgrounds, fit, shapes.geometric, decorations.pathreplacing,arrows.meta}

\ifCLASSINFOpdf
\else
\fi

\begin{document}

\title{\huge Reliability-Constrained Blind Beam Alignment for Backscatter-MIMO mounted Target in Cluttered Multipath Channels}
%\thanks{National Foundation (NSFC), NO.12141107 supports this work.}}

\author{
Xuehui~Dong,~\IEEEmembership{Student Member,~IEEE,}
Kai~Wan,~\IEEEmembership{Member,~IEEE,} 
Gui~Zhou,~\IEEEmembership{Member,~IEEE,} 
Chen Shao,
Miyu Feng,
and~Robert~Caiming~Qiu,~\IEEEmembership{Fellow,~IEEE}

\thanks{The authors are with the School of Electronic Information and Communications, Huazhong University of Science and Technology, Wuhan 430074, China (e-mail: \{xuehuidong,  kai\_wan, gui\_zhou, shaochen0517,  miyu\_feng, caiming\}@hust.edu.cn).}
}

\maketitle

\begin{abstract}
Practical integrated sensing and communication (ISAC) is challenged by strong static clutter and dense non-line-of-sight (NLoS) multipath, which bury target-coupled echoes and create spurious spatial peaks for beam alignment. Most receiver-side countermeasures still treat the target as an electromagnetically passive scatterer, limiting their ability to make the target echo structurally distinguishable from the environment. This paper reveals a structural correspondence between these barriers and target-side Backscatter-MIMO responses: reflection modulation separates the target echo from unmodulated clutter in the waveform domain, while retro-directional passive beamforming concentrates the tagged echo toward the BS-facing direction and mitigates NLoS-induced false-peak locking. To turn this correspondence into a usable ISAC link, reliable dual-end spatial locking is required to close the cascaded backscatter link budget and provide beam-domain angular information. We propose a downlink-triggered blind dual-end alignment protocol that selects both BS and Backscatter-MIMO codeword indices from the tagged echo observed at the BS, without active pilots, CSI feedback, or global synchronization at the target. We further derive a clutter-aware remodulation waveform robust to fractional timing offsets and construct adjustable-width BS/Backscatter-MIMO codebooks using quadratic phase spoiling. The reliability analysis gives closed-form expressions for the coherence-averaged end-to-end success probability. It shows that narrower beams are not always better: in NLoS-dominated regimes, increasing the array size can even degrade alignment reliability. The optimal beamwidth is instead governed by the cross-phase competition between discovery and alignment, yielding a nontrivial feasible region whose boundary is characterized by the derived expressions. Simulations validate the analysis and show that the proposed framework improves reliability-gated locked-link performance under strong clutter, severe NLoS multipath, and finite coherence time.
\end{abstract}


\IEEEpeerreviewmaketitle

\section{Introduction}
\label{sec:introduction}
Integrated sensing and communication (ISAC) is increasingly regarded as a key capability of future wireless networks, where a common radio platform supports data delivery, environmental awareness, and target sensing~\cite{liu2022integrated,xiong2024torch}. In directional mmWave systems, angular acquisition and beam locking are especially fundamental because localization, tracking, and high-gain communication all depend on fast and reliable identification of the correct spatial beams~\cite{xiao2017millimeter}. Existing beam-management studies have made substantial progress through hierarchical search, sparse or compressed-sensing channel estimation, and sensing-assisted beam prediction that uses radar or multimodal observations to reduce training overhead~\cite{alkhateeb2014channel,gao2025integrated}. These methods improve receiver-side inference, but the object of interest remains an electromagnetically passive scatterer. As a result, the target echo itself carries no controllable waveform or spatial signature that would allow structural discrimination from clutter and multipath.

Two physical impairments become dominant in practical cluttered deployments. First, static environmental clutter generated by buildings, vehicles, metallic infrastructure, and other stationary reflectors can overwhelm the weak target-coupled echo, often driving the signal-to-clutter ratio far below unity~\cite{richards2014fundamentals,wang2024clutter}. For stationary or slowly moving targets, Doppler filtering offers limited relief because zero-Doppler target returns are spectrally indistinguishable from static clutter. Second, rich non-line-of-sight (NLoS) multipath can create virtual angular peaks during beam sweeping, so that the strongest return may correspond to an indirect path rather than the desired target-related direction~\cite{alkhateeb2014channel}. Existing remedies, ranging from space-time adaptive processing and clutter-aware waveform or precoder design to interference-aware signaling and resource allocation, have improved ISAC robustness from the receiver and network-control sides~\cite{ward1994space,liu2018toward,yuan2021integrated,niu2024interference,wang2024clutter}. However, they still operate under the passive-target premise, where clutter is estimated and cancelled after reception and multipath is modeled or suppressed after it has already distorted the beam search.

Programmable surfaces and backscatter architectures provide the closest route to relaxing this premise, but existing studies have not yet addressed the structural gap considered here. RIS-aided wireless systems use controllable passive elements to reshape propagation, enhance coverage, and create virtual LoS paths for communication, sensing, and localization~\cite{wu2019towards,di2020smart,zhang2021enabling,zhou2020framework}. Ambient backscatter and symbiotic radio exploit reflection modulation for ultra-low-power communication, while recent backscatter-ISAC and retro-directive backscatter studies further explore sensing, localization, and array-enabled passive links~\cite{liu2013ambient,van2018ambient,liang2020symbiotic,he2020monostatic,lotti2025realtime,zhao2024bisac,dong2025metasurface}. Nevertheless, RIS is still mainly used as a channel-enhancement tool, and backscatter devices are usually treated as communication nodes or auxiliary anchors under pilot, CSI, or synchronization assumptions. To the best of our knowledge, few prior work has exploited the intrinsic Backscatter-MIMO correspondence between reflection modulation and clutter separation, or between retro-directional passive beamforming and NLoS false-peak suppression, under standalone constraints. This motivates a shift from receiver-side interference rejection to target-side electromagnetic response design.

\begin{figure*}[t]
    \centering
    \includegraphics[width=0.76\linewidth]{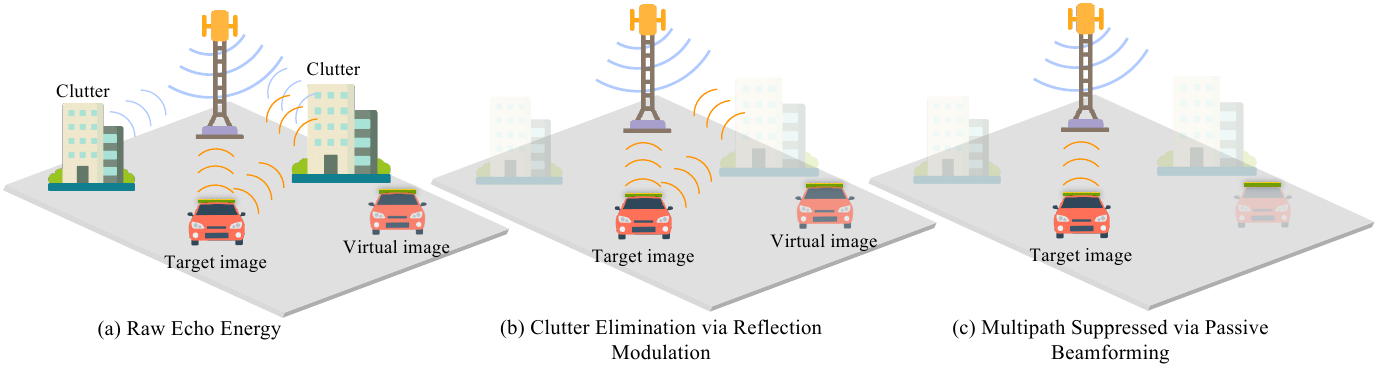}
    \caption{Core mechanisms of the proposed standalone Backscatter-MIMO cooperative target sensing framework.}
    \label{fig:core_mechanisms}
\end{figure*}

This paper identifies a different perspective. As illustrated in Fig.~\ref{fig:core_mechanisms}, a target equipped with a Backscatter-MIMO array is no longer electromagnetically passive. Two intrinsic attributes of Backscatter-MIMO form a structural correspondence with the clutter and multipath barriers. First, reflection modulation embeds a controllable signature into the backscattered waveform, so that the tagged echo occupies a waveform-domain subspace distinguishable from unmodulated static clutter, as shown in Fig.~\ref{fig:core_mechanisms}(b). Hence, clutter suppression is transformed from estimating an unknown interference component into separating a structured tagged echo from the raw return in Fig.~\ref{fig:core_mechanisms}(a). Second, retro-directional passive beamforming shapes the reflected wavefront toward the BS-facing direction, making the desired tagged echo geometrically distinct from diffuse NLoS scattering, as shown in Fig.~\ref{fig:core_mechanisms}(c). Thus, multipath suppression becomes a geometry-aware focusing problem rather than purely receiver-side filtering. Importantly, both mechanisms require no active RF chains, dedicated pilots, or CSI feedback at the target side. To the best of our knowledge, this structural correspondence has not been systematically exploited for blind dual-end alignment of standalone cooperative targets in cluttered multipath ISAC.

The benefits of these mechanisms cannot be realized without \emph{reliable beam alignment}. From the communication perspective, Backscatter-MIMO suffers from severe cascaded round-trip path loss, and dual-end array gain is essential to close the link budget. From the sensing perspective, the selected beam index provides an initial beam-domain angular observation, while the post-lock beamwidth determines the residual angular ambiguity of the locked spatial link. Therefore, beam alignment is not merely a communication-training step; it is the mechanism that converts the controllable Backscatter-MIMO response into usable sensing and communication performance. However, standalone Backscatter-MIMO makes this alignment problem nontrivial: the target transmits no active uplink pilots, estimates no CSI, and shares no global clock with the BS, so all beam decisions must be inferred from the tagged echo observed at the BS side. Moreover, fractional asynchronous timing offsets distort the effective tag waveform and weaken matched-filter detection~\cite{zhao2024performance, huang2025riding}, and the entire discovery-and-alignment procedure must complete within a finite channel-coherent operation window before the acquired spatial information becomes outdated. These constraints raise the central question of this paper: \emph{how can a standalone Backscatter-MIMO target be reliably discovered and dual-end aligned in cluttered multipath channels without active feedback, CSI exchange, or global synchronization?}

To address this challenge, this paper proposes a downlink-triggered blind dual-end beam alignment protocol for standalone Backscatter-MIMO in cluttered multipath channels. The protocol leverages reflection modulation for waveform-domain clutter separation and retro-directional passive beamforming for mitigating NLoS-induced false-peak locking, and completes dual-end beam locking solely from the tagged echo observed at the BS under the standalone target constraint. The design objective is not to maximize a pre-alignment received power or an instantaneous rate, but to establish a reliable high-SNR, low-ambiguity locked link within the available coherence window. The main contributions of this paper are summarized as follows.

\begin{itemize}
    \item To the best of our knowledge, this is the first work to establish a cooperative target-sensing paradigm in which the target-mounted Backscatter-MIMO array programmably reshapes its own electromagnetic response to exploit a structural correspondence between cluttered-multipath impairments and target-side controllable attributes. Reflection modulation makes the target echo separable from unmodulated clutter in the waveform domain, while retro-directional passive beamforming concentrates the tagged echo geometrically away from diffuse NLoS scattering in the spatial domain.

    \item We develop a downlink-triggered blind dual-end alignment protocol for standalone Backscatter-MIMO, in which target discovery reuses regular downlink transmissions without dedicated sensing resources. Upon discovery, the protocol performs BS transmit-codebook sweeping, Backscatter-MIMO reflection-codebook sweeping, and beam-index-based locking, with all decisions made solely from the tagged echo at the BS, without active pilots, CSI feedback, or global synchronization at the target.

    \item We derive a closed-form remodulation waveform and adjustable-width dual-end codebooks. The waveform resolves the tension between clutter-subspace orthogonality and fractional-timing-offset robustness via a projected Rayleigh quotient maximization. The Backscatter-MIMO reflection codebook is designed retro-directionally to uniquely map each incident direction to a single dominant reflected path, enabling unambiguous passive direction locking.

    \item We derive tractable reliability expressions for the single-frame discovery probability, the incident-beam alignment outage probability, and the coherence-averaged end-to-end success probability. The analysis shows that narrower beams are not always better: in NLoS-dominated regimes, increasing the array size can degrade alignment reliability. It further reveals the cross-phase competition between discovery and alignment beamwidths, and characterizes a nontrivial reliability-feasible region whose boundary determines the narrowest reliability-feasible BS sweeping beamwidth.
\end{itemize}

The remainder of this paper is organized as follows. Section~\ref{sec:system_model} presents the clutter-aware channel, asynchronous remodulation, and beamwidth-parameterized echo models. Section~\ref{sec:protocol_problem} introduces the proposed protocol and reliability-constrained problem formulation. Section~\ref{sec:codebook_waveform_design} develops the codebook and waveform designs. Section~\ref{sec:performance} analyzes discovery, alignment, and end-to-end reliability. Section~\ref{sec:simulation} provides numerical results.  Section~\ref{sec:conclusion} concludes the paper.

\section{System Model}
\label{sec:system_model}

\subsection{Clutter-Aware Backscatter-MIMO Channel}
\label{subsec:channel_model}

We consider a monostatic ISAC BS equipped with an $M_{\mathrm{ant}}$-element ULA, which aims to discover and lock a standalone Backscatter-MIMO target equipped with an $N$-element reconfigurable backscatter ULA. The BS element spacing is set to $d_{\mathrm{BS}}=\lambda/2$, while the Backscatter-MIMO element spacing is set to $d_{\mathrm{BM}}=\lambda/4$ to avoid grating-lobe ambiguity in the round-trip reflection response. 
For notational brevity, we write $\mathbf{a}_{M_{\mathrm{ant}}}(\theta)=\mathbf{a}_{M_{\mathrm{ant}}}(\theta;d_{\mathrm{BS}})$ and $\mathbf{a}_{N}(\theta)=\mathbf{a}_{N}(\theta;d_{\mathrm{BM}})$ in the sequel.
\begin{figure}[ht]
\centering
\includegraphics[width=0.75\linewidth]{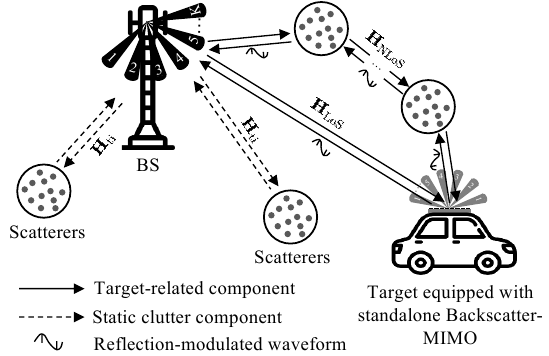}
\caption{ The system model of beam alignment between the BS and the cooperative target equipped with Backscatter-MIMO.}
\label{fig:monostatic_channel}
\end{figure}

As illustrated in Fig.~\ref{fig:monostatic_channel}, 
the one-way channel from the BS to the Backscatter-MIMO target is modeled as a geometry-based Rician channel,
\begin{equation}\small
\mathbf{H}_{\mathrm{tv}}=\sqrt{\frac{\kappa}{\kappa+1}}\mathbf{H}_{\mathrm{LoS}}+\sqrt{\frac{1}{\kappa+1}}\mathbf{H}_{\mathrm{NLoS}},
\label{eq:H_tv}
\end{equation}
where $\kappa$ is the Rician factor. The deterministic LoS component is
$\mathbf{H}_{\mathrm{LoS}}=\beta_{0}\mathbf{a}_{N}(\theta_{\mathrm{rx},0})\mathbf{a}_{M_{\mathrm{ant}}}^{\top}(\theta_{\mathrm{tx},0})$,
where $\beta_{0}\in\mathbb{C}$ is the one-way complex gain of the desired target-coupled path, and $\theta_{\mathrm{tx},0}$ and $\theta_{\mathrm{rx},0}$ denote the LoS AoD at the BS and AoA at the Backscatter-MIMO, respectively. The NLoS component consists of $P$ scattering paths,
\begin{equation}\small
\mathbf{H}_{\mathrm{NLoS}}=\frac{1}{\sqrt{P}}\sum_{p=1}^{P}\beta_{p}\mathbf{a}_{N}(\theta_{\mathrm{rx},p})\mathbf{a}_{M_{\mathrm{ant}}}^{\top}(\theta_{\mathrm{tx},p}),
\label{eq:H_NLoS}
\end{equation}
where $\beta_{p}\in\mathbb{C}$ is the one-way complex gain of the $p$-th NLoS path, and $(\theta_{\mathrm{tx},p},\theta_{\mathrm{rx},p})$ are the corresponding AoD and AoA pair. These NLoS paths may create false spatial energy peaks during beam sweeping.

In addition to the target-coupled channel, the BS also receives strong static environmental clutter. Let $Q$ dominant static scatterers be located at angles $\{\phi_q\}_{q=1}^{Q}$. Their spatial response matrix is written as
\begin{equation}\small
\mathbf{H}_{\mathrm{ti}}=\left[\mathbf{a}_{M_{\mathrm{ant}}}(\phi_1),\mathbf{a}_{M_{\mathrm{ant}}}(\phi_2),\ldots,\mathbf{a}_{M_{\mathrm{ant}}}(\phi_Q)\right]^{\top},
\label{eq:H_ti}
\end{equation}
and the corresponding round-trip gain matrix is
$\boldsymbol{\Lambda}=\mathrm{diag}\{\lambda_1^2,\lambda_2^2,\ldots,\lambda_Q^2\}$,
where $\lambda_q\in\mathbb{C}$ denotes the one-way complex gain associated with the $q$-th static scatterer. The static-clutter round-trip channel is therefore
\begin{equation}\small
\mathbf{H}_{\mathrm{sc}}=\mathbf{H}_{\mathrm{ti}}^{\top}\boldsymbol{\Lambda}\mathbf{H}_{\mathrm{ti}}\in\mathbb{C}^{M_{\mathrm{ant}}\times M_{\mathrm{ant}}}.
\label{eq:H_sc}
\end{equation}
Under this convention, both the target-coupled echo and the static-clutter echo are represented by squared one-way gains in the round-trip channel. The instantaneous signal-to-clutter ratio (SCR) of the Backscatter-MIMO LoS echo relative to the aggregate static clutter is defined as
$\mathrm{SCR}\triangleq\frac{|\beta_0|^4}{\sum_{q=1}^{Q}|\lambda_q|^4}$.
In clutter-dominated environments, $\mathrm{SCR}\ll1$, which makes the natural target echo difficult to distinguish from static environmental reflections.

\subsection{Fractionally Asynchronous Remodulation Model}
\label{subsec:async_model}

A standalone Backscatter-MIMO target is not globally synchronized with the BS. Therefore, the remodulation waveform imposed by the target may arrive at the BS with a fractional sampling offset. 

We sample the received echo at discrete time instances $t = nT_s$, where $T_s$ is the sampling interval and $n = 0, 1, \ldots, N_s - 1$, with $N_s$ being the total number of samples collected during the observation period. Let the baseband transmitted symbol sequence be $\mathbf{s} = [s_0, s_1, \ldots, s_{N_s-1}]^\top$. The continuous-time transmitted signal is generated by passing these symbols through a transmit pulse-shaping filter $g_{\mathrm{tx}}(t)$, i.e., $x(t) = \sum_{n=0}^{N_s-1} s_n g_{\mathrm{tx}}(t - nT_s)$.

At the Backscatter-MIMO side, the reflection modulation embeds a specific tag sequence $\boldsymbol{\alpha} = [\alpha[0], \alpha[1], \dots, \alpha[N_s-1]]^\top$ into the backscattered waveform. To ensure reliable detection regardless of the arbitrary block arrival time, the Backscatter-MIMO continuously applies the sequence in a cyclic manner, yielding the cyclic remodulation waveform $\alpha_{\mathrm{cyc}}(t) = \sum_{m=-\infty}^{+\infty} \sum_{n=0}^{N_s-1} \alpha[n]\, g_{\mathrm{bsm}}(t - nT_s - mN_sT_s)$, where $g_{\mathrm{bsm}}(t)$ denotes the equivalent pulse-shaping filter of the Backscatter-MIMO impedance switching waveform.

Due to the absence of a shared clock between the BS and the standalone Backscatter-MIMO, the reflection modulation is fundamentally asynchronous. We denote the relative timing offset as $T_{\mathrm{asyn}} = (k + \Delta\tau)T_s$, where $k \in \mathbb{Z}$ represents the integer symbol delay and $\Delta\tau \in [0,1)$ is the fractional delay offset. Consequently, the backscattered signal over the observation window $t \in [0, N_s T_s)$ is modulated by the shifted cyclic tag waveform, i.e., $x_{\mathrm{refl}}(t) = x(t) \cdot \alpha_{\mathrm{cyc}}(t - T_{\mathrm{asyn}})$.

At the receiver, the effective sampled tag sequence $\tilde{\alpha}[n] \triangleq \alpha_{\mathrm{cyc}}(nT_s - T_{\mathrm{asyn}})$ inherently suffers from ISI due to the fractional delay $\Delta\tau$. By letting $i = n - k - m$, this sampling process expands as:
\begin{equation}\small
\begin{aligned}
    \tilde{\alpha}[n] = &\sum_{i=-\infty}^{+\infty} \alpha[\langle n-k-i \rangle_{N_s}]\, g_{\mathrm{bsm}}((i - \Delta\tau)T_s) \\
    \approx &\alpha[\langle n-k \rangle_{N_s}]\, g_{\mathrm{bsm}}(-\Delta\tau T_s) \\
    &+ \alpha[\langle n-k-1 \rangle_{N_s}]\, g_{\mathrm{bsm}}((1-\Delta\tau)T_s),
\end{aligned}
\end{equation}
where $\langle \cdot \rangle_{N_s}$ denotes the modulo-$N_s$ operation, and the approximation holds under the practical assumption that the energy of $g_{\mathrm{bsm}}(t)$ is primarily concentrated within adjacent symbol periods.

\subsection{Beamwidth-Parameterized Echo Model}
\label{subsec:beamwidth_echo_model}

The proposed protocol uses two beamwidth parameters. The default Backscatter-MIMO beamwidth $\Theta_{\mathrm{d}}$ specifies the broad-in/out reflection mode used during downlink-assisted discovery and BS-side transmit sweeping. The BS alignment beamwidth $\Theta_{\mathrm{a}}$ specifies the transmit sweeping codebook used after the discovery event has been triggered. For a prescribed $\Theta_{\mathrm{a}}$, the BS transmit codebook is denoted by
$\mathcal{W}(\Theta_{\mathrm{a}})=\left\{\mathbf{w}_{k}(\Theta_{\mathrm{a}})\right\}_{k=1}^{K(\Theta_{\mathrm{a}})}$,
where $K(\Theta_{\mathrm{a}})$ is the number of BS transmit beams required to cover the angular search region. For a prescribed Backscatter-MIMO reflection beamwidth $\Theta$, the reflection codebook is denoted by
$\mathcal{F}(\Theta)=\left\{\boldsymbol{\Phi}_{l}(\Theta)\right\}_{l=1}^{L(\Theta)}$,
where $L(\Theta)$ is the number of passive reflection configurations.

For a generic BS transmit beam $\mathbf{w}$ and Backscatter-MIMO reflection configuration $\boldsymbol{\Phi}(\Theta)$, the effective round-trip Backscatter-MIMO channel is
\begin{equation}\small
\mathbf{H}_{\mathrm{bsm}}(\Theta)=\mathbf{H}_{\mathrm{tv}}^{\top}\boldsymbol{\Phi}(\Theta)\mathbf{H}_{\mathrm{tv}}\in\mathbb{C}^{M_{\mathrm{ant}}\times M_{\mathrm{ant}}}.
\label{eq:H_bsm}
\end{equation}
Let $x(t)$ denote the BS transmitted baseband waveform with average power $\mathbb{E}\{|x(t)|^2\}=P_{\mathrm{tx}}$, and let $\alpha(t)$ denote the continuous-time remodulation signature generated from $\boldsymbol{\alpha}$. The received echo at the BS is modeled as
\begin{equation}\small
\mathbf{y}(t)=\mathbf{H}_{\mathrm{bsm}}(\Theta)\mathbf{w}\alpha(t)x(t)+\mathbf{H}_{\mathrm{sc}}\mathbf{w}x(t)+\mathbf{n}(t),
\label{eq:received_echo}
\end{equation}
where $\mathbf{n}(t)\sim\mathcal{CN}(\mathbf{0},\sigma^2\mathbf{I}_{M_{\mathrm{ant}}})$ is the receiver noise. The first term is the tagged Backscatter-MIMO echo, while the second term is the unmodulated static-clutter echo. In Phase-I, $\Theta=\Theta_{\mathrm{d}}$ and $\mathbf{w}$ corresponds to the ongoing downlink transmission. In Phase-II, $\Theta=\Theta_{\mathrm{d}}$ and $\mathbf{w}$ is swept over $\mathcal{W}(\Theta_{\mathrm{a}})$. In Phase-III, the BS beam is locked to the selected $\mathbf{w}_{k^{\star}}(\Theta_{\mathrm{a}})$, while the Backscatter-MIMO sweeps the narrowest retro-directional reflection codebook.

By collecting the effective samples into $\tilde{\boldsymbol{\alpha}} = [\tilde{\alpha}[0], \tilde{\alpha}[1], \dots, \tilde{\alpha}[N_s-1]]^\top$, the discrete-time received signal matrix $\mathbf{Y} \in \mathbb{C}^{M_{\mathrm{ant}} \times N_s}$ over the observation period is:
\begin{equation}\small
    \mathbf{Y} = \mathbf{H}_{\mathrm{bsm}}(\Theta)\, \mathbf{w}(\Theta)\, (\tilde{\boldsymbol{\alpha}} \circ \mathbf{x})^\top + \mathbf{H}_{\mathrm{sc}}\, \mathbf{w}(\Theta)\, \mathbf{x}^\top + \mathbf{N},
    \label{eq:Y_discrete}
\end{equation}
where $\mathbf{x} = [x[0], x[1], \ldots, x[N_s-1]]^\top \in \mathbb{C}^{N_s \times 1}$ is the sampled transmitted signal vector, $\circ$ denotes the Hadamard product, and $\mathbf{N} \in \mathbb{C}^{M_{\mathrm{ant}} \times N_s}$ is the AWGN matrix. The first term carries the modulated Backscatter-MIMO echo, where the Hadamard product $\tilde{\boldsymbol{\alpha}} \circ \mathbf{x}$ captures the joint effect of the remodulation signature and the fractional ISI distortion. The second term represents the unmodulated static clutter, whose baseband envelope is strictly proportional to $\mathbf{x}$ and thus occupies a distinct waveform subspace from the modulated echo, forming the physical basis for clutter separation via matched filtering.

After dual-end beam locking, the link can be evaluated by standard communication and sensing quantities. For backscatter communication, the locked-link SNR is determined by the BS transmit gain and the Backscatter-MIMO reflection gain. With $G_{\mathrm{BS}}(\Theta_{\mathrm a})\approx 2/\Theta_{\mathrm a}$ and $G_{\mathrm{BM}}(\Theta_{\mathrm{BM}})\approx 2/\Theta_{\mathrm{BM}}$, it is approximated as
\begin{equation}\small
\rho_{\mathrm{lock}}(\Theta_{\mathrm a},\Theta_{\mathrm{BM}})\approx \frac{4P_{\mathrm{tx}}|\beta_0|^4}{\sigma^2\Theta_{\mathrm a}\Theta_{\mathrm{BM}}},
\label{eq:rho_lock}
\end{equation}
where $\Theta_{\mathrm a}$ and $\Theta_{\mathrm{BM}}$ denote the locked BS beamwidth and Backscatter-MIMO reflection beamwidth, respectively. The corresponding locked-link spectral efficiency $R_{\mathrm{lock}}=\log_2(1+\rho_{\mathrm{lock}})$ is monotonic in $\rho_{\mathrm{lock}}$.
For angular sensing, the selected BS beam localizes the target within a beam bin of width $\Theta_{\mathrm a}$, yielding the beam-domain ambiguity $\sigma_{\theta,\mathrm{lock}}^2(\Theta_{\mathrm a})\approx\Theta_{\mathrm a}^2/12$ under a uniform residual uncertainty model. Hence, reducing $\Theta_{\mathrm a}$ simultaneously increases the locked-link SNR and reduces the angular ambiguity.

\section{Protocol-Guided Problem Formulation}
\label{sec:protocol_problem}
This section formulates standalone Backscatter-MIMO alignment as a reliability-constrained link-locking problem. The default Backscatter-MIMO beamwidth $\Theta_{\mathrm d}$ governs opportunistic discovery, while the BS alignment beamwidth $\Theta_{\mathrm a}$ governs post-trigger sweeping latency and locked-link quality. Using the coherence time $T_{\mathrm{coh}}\sim\mathrm{Exp}(\mu)$, frame duration $T_{\mathrm{frame}}$, dwell time $T_{\mathrm{dwell}}$, reflection-sweeping duration $T_{\mathrm{III}}$, and feedback overhead $T_{\mathrm{fb}}$, we define the parent problem and its codebook/waveform design subproblems.

\subsection{Event-Triggered Discovery-and-Locking Protocol}
\label{subsec:protocol}
% -------------------------------------------------------------------------

\begin{figure}[htbp]
    \centering
    \resizebox{\columnwidth}{!}{%
    \begin{tikzpicture}[
        >=Latex,
        font=\rmfamily\small, % 提升基础字号，避免缩放后过小
        line cap=round,
        line join=round,
        lifeline/.style={black!55, line width=0.75pt},
        phaseA/.style={fill=black!2.5, rounded corners=2pt},
        phaseB/.style={fill=black!5.5, rounded corners=2pt},
        header/.style={draw=black!75, line width=0.6pt, fill=white, rounded corners=2pt, font=\rmfamily\bfseries\small, align=center, inner xsep=4pt, inner ysep=3pt},
        prep/.style={draw=black!70, line width=0.65pt, fill=black!1.5, rounded corners=3pt, align=center, text width=11.2cm, inner sep=2.5mm}, % 增大宽度，完美放下一行
        bsbox/.style={draw=black!65, line width=0.55pt, fill=white, rounded corners=2pt, align=center, text width=5cm, inner sep=1.5mm}, % 放宽，避免文本溢出
        tagbox/.style={draw=black!65, line width=0.55pt, fill=white, rounded corners=2pt, align=center, text width=3.6cm, inner sep=1.5mm}, % 放宽，避免 remodulation 被切断
        phaseLab/.style={font=\rmfamily\bfseries\normalsize, align=center, text width=3.4cm},
        msg/.style={->, black, line width=0.65pt},
        echo/.style={->, black, line width=0.65pt, dashed, dash pattern=on 3pt off 2pt},
        aux/.style={->, black!45, line width=0.55pt, dashed, dash pattern=on 3pt off 3pt},
        note/.style={font=\rmfamily\scriptsize, inner sep=1pt, text opacity=1},
        auxnote/.style={font=\rmfamily\itshape\scriptsize, text=black!55, inner sep=1pt, text opacity=1}
    ]

    % ==================================================
    % 绝对坐标系 (宽敞布局)
    % ==================================================
    \def\xL{-5.0}     % 背景左边缘
    \def\xPhase{-3.4} % 阶段标签中心
    \def\xBS{0}       % 基站生命线
    \def\xTAG{5.5}    % 标签生命线
    \def\xR{7.6}      % 背景右边缘
    \def\bottom{-14.8}% 底部边界

    % ==================================================
    % Offline preparation
    % ==================================================
    \node[prep] at (1.30, 0.70) {
        {\bfseries\large Offline waveform/codebook preparation}\\[1.5mm]
        $\bullet$ Design $\boldsymbol{\alpha}^{*}$ \hfill
        $\bullet$ Construct $\mathcal{W}(\Theta_{\mathrm{a}})$ \hfill
        $\bullet$ Construct $\mathcal{F}(\Theta)$ \hfill
        $\bullet$ Select $\Theta_{\mathrm{d}},\Theta_{\mathrm{a}}$
    };

    % ==================================================
    % Column headers and lifelines
    % ==================================================
    \draw[lifeline] (\xBS,-0.75) -- (\xBS,\bottom);
    \draw[lifeline] (\xTAG,-0.75) -- (\xTAG,\bottom);
    
    \node[header, minimum width=2.0cm] at (\xBS, -0.60) {Base station};
    \node[header, minimum width=2.8cm] at (\xTAG, -0.60) {Backscatter-MIMO};
    % ==================================================
    % Phase backgrounds
    % ==================================================
    \begin{scope}[on background layer]
        \fill[phaseA] (\xL,-1.0) rectangle (\xR,-4.3);
        \fill[phaseB] (\xL,-4.4) rectangle (\xR,-8.3);
        \fill[phaseA] (\xL,-8.4) rectangle (\xR,-11.7);
        \fill[phaseB] (\xL,-11.8) rectangle (\xR,-14.8);
    \end{scope}

    % ==================================================
    % Phase labels
    % ==================================================
    \node[phaseLab] at (\xPhase,-2.2) {Phase-I:\\Downlink-assisted\\target discovery};
    \node[phaseLab] at (\xPhase,-6.2) {Phase-II:\\BS transmit\\beam sweeping};
    \node[phaseLab] at (\xPhase,-9.6) {Phase-III:\\Backscatter-MIMO\\retro-directional\\pattern sweeping};
    \node[phaseLab] at (\xPhase,-13.2) {Phase-IV:\\Alignment\\locking\\\& Operation};

    % ==================================================
    % Phase I: Downlink-assisted target discovery
    % ==================================================
    \node[tagbox] at (\xTAG,-1.6)
        {Default broad mode\\$\boldsymbol{\Phi}(\Theta_{\mathrm{d}})+\boldsymbol{\alpha}^{*}$-remodulation};

    \draw[msg] (\xBS,-2.5) -- (\xTAG,-2.5)
        node[midway, above, note] {Ongoing downlink signal $\mathbf{x}$};

    \draw[echo] (\xTAG,-2.8) -- (\xBS,-2.8)
        node[midway, below, note] {Asynchronous echo $\tilde{\boldsymbol{\alpha}}\circ\mathbf{x}$ + clutter};

    \node[bsbox] at (\xBS,-3.7)
        {MF + CFAR detector $(P_D(\Theta_{\mathrm{d}}))$\\\& Trigger beam alignment};

    % ==================================================
    % Phase II: BS transmit beam sweeping
    % ==================================================
    \node[bsbox] at (\xBS,-4.9)
        {Tx beam sweeping $\mathbf{w}_k\in\mathcal{W}(\Theta_{\mathrm{a}})$};

    \node[tagbox] at (\xTAG,-4.9)
        {Maintains default mode\\$\boldsymbol{\Phi}(\Theta_{\mathrm{d}})+\boldsymbol{\alpha}^{*}$-remodulation};

    \draw[msg] (\xBS,-5.8) -- (\xTAG,-5.8)
        node[midway, above, note] {Dedicated CW beam $(k=1\cdots K)$};

    % Clutter (确保不碰到左侧标签)
    \draw[aux] (\xBS,-5.8) .. controls (-1.2,-6.2) and (-1.2,-6.4) .. (\xBS,-6.8);
    \node[auxnote] at (-0.6,-6.8) {Clutter};

    % Multiple paths
    \draw[aux] (\xTAG,-6.8) .. controls (3.5,-6.1) and (1.5,-6.4) .. (\xBS,-6.8);
    \node[auxnote] at (2.75,-6.2) {Multiple paths};

    \draw[echo] (\xTAG,-6.8) -- (\xBS,-6.8)
        node[midway, below, note] {Tagged echo under CW probing};

    \node[bsbox] at (\xBS,-7.7)
        {Determine optimal transmit index $k^{*}$\\BS-side alignment $(P_{\mathrm{out}}(\Theta_{\mathrm{a}},\Theta_{\mathrm{d}}))$};

    % ==================================================
    % Phase III: Backscatter-MIMO retro-directional pattern sweeping
    % ==================================================
    \draw[msg] (\xBS,-9.0) -- (\xTAG,-9.0)
        node[midway, above, note] {DCI notification $(k^{*})$ + locked CW beam $\mathbf{w}_{k^{*}}$};

    \node[tagbox] at (\xTAG,-9.7)
        {Reflection sweep mode\\$\boldsymbol{\Phi}_l\in\mathcal{F}(\Theta_{\min})$};

    % Multiple paths
    \draw[aux] (\xTAG,-10.6) .. controls (3.5,-10.0) and (1.5,-10.0) .. (\xBS,-10.6);
    \node[auxnote] at (2.75,-10.32) {Multiple paths};

    \draw[echo] (\xTAG,-10.6) -- (\xBS,-10.6)
        node[midway, below, note] {Tagged echoes under $\boldsymbol{\Phi}_1\cdots\boldsymbol{\Phi}_L$};

    \node[bsbox] at (\xBS,-11.3)
        {Determine optimal reflection index $l^{*}$};

    % ==================================================
    % Phase IV: Alignment locking and operation
    % ==================================================
    \draw[msg] (\xBS,-12.3) -- (\xTAG,-12.3)
        node[midway, above, note] {DCI feedback $(k^{*},l^{*})$};

    \node[tagbox] at (\xTAG,-13.0)
        {Lock reflection\\configuration $\boldsymbol{\Phi}_{l^{*}}(\Theta_{\min})$};

    \draw[<->, black, line width=0.75pt] (\xBS,-14.0) -- (\xTAG,-14.0)
        node[midway, above, note, font=\rmfamily\bfseries\small] {Sensing or data transmission};

    \node[note, font=\rmfamily\bfseries\small] at (2.75,-14.4)
        {Protocol-level success probability $P_{\mathrm{e2e}}$};

    \end{tikzpicture}%
    }
    \caption{Protocol flow of the proposed Backscatter-MIMO beam alignment and operation procedure.}
    \label{fig:protocol_flow}
\end{figure}
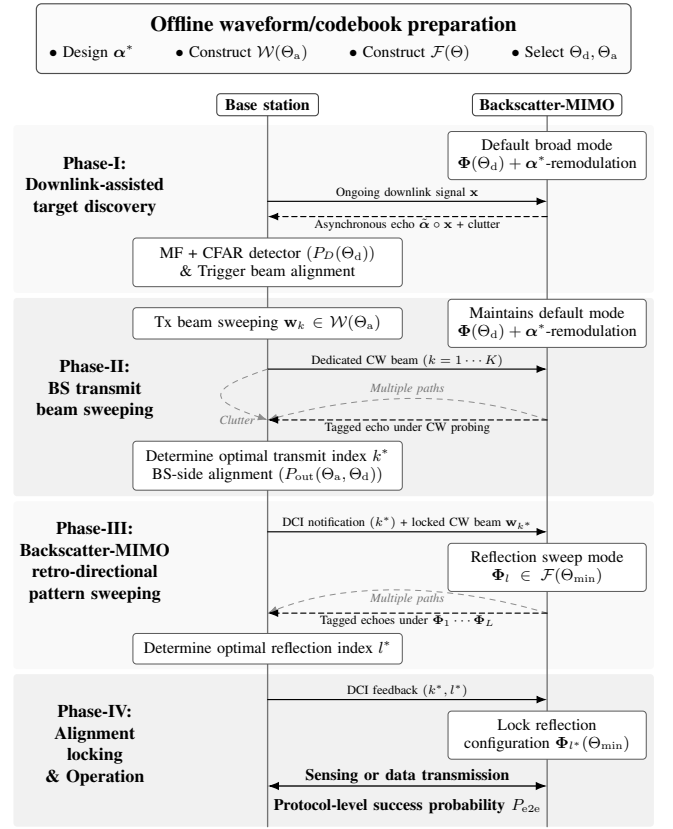

The proposed protocol aims to establish a high-quality locked spatial link without active uplink pilots, CSI feedback, or global synchronization at the Backscatter-MIMO target. The BS infers target existence and direction only from echoes induced by its downlink or probing signals. As shown in Fig.~\ref{fig:protocol_flow}, the protocol consists of one offline preparation stage and four online phases.

\noindent\textbf{Offline preparation.}
The BS and the Backscatter-MIMO target predefine the remodulation waveform $\boldsymbol{\alpha}$, the BS transmit codebook $\mathcal{W}(\Theta_{\mathrm a})$, and the Backscatter-MIMO reflection codebook $\mathcal{F}(\Theta)$, which provide tagged-echo extraction and controllable angular coverage/gain.

\noindent\textbf{Phase-I: Downlink-assisted target discovery.}
The BS reuses ongoing downlink transmissions as opportunistic illumination. The target stays in the default broad-in/out mode $\boldsymbol{\Phi}(\Theta_{\mathrm d})$, remodulates the incident waveform with $\boldsymbol{\alpha}$, and triggers a discovery event $\mathcal{D}_{\Theta_{\mathrm d}}$ through CFAR detection at the BS.

\noindent\textbf{Phase-II: BS transmit beam sweeping.}
After discovery, the BS sweeps $\mathcal{W}(\Theta_{\mathrm a})$ with dedicated probing, while the target keeps $\boldsymbol{\Phi}(\Theta_{\mathrm d})$. The incident beam is selected as
\begin{equation}\small
k^{\star}=\arg\max_{1\leq k\leq K(\Theta_{\mathrm a})}\mathcal{M}_{k},
\label{eq:k_star_protocol}
\end{equation}
where $\mathcal{M}_{k}$ is the tagged-echo metric for the $k$-th BS beam. With $k_0$ denoting the desired LoS incident-beam index, the dominant pre-lock alignment error is $k^{\star}\neq k_0$, mainly caused by NLoS-induced false peaks.

\noindent\textbf{Phase-III: Backscatter-MIMO reflection-pattern sweeping.}
The BS locks to $\mathbf{w}_{k^{\star}}(\Theta_{\mathrm a})$, and the target sweeps the narrowest retro-directional codebook $\mathcal{F}(\Theta_{\min}^{\mathrm{BM}})$. The reflection state is selected as
\begin{equation}\small
l^{\star}=\arg\max_{1\leq l\leq L(\Theta_{\min}^{\mathrm{BM}})}\mathcal{Q}_{l},
\label{eq:l_star_protocol}
\end{equation}
where $\mathcal{Q}_{l}$ is the tagged-echo metric for the $l$-th reflection configuration. Phase-III is treated as passive reflection refinement: its time cost is included in the post-trigger duration, while the dominant reliability loss is captured by $k^{\star}\neq k_0$.

\noindent\textbf{Phase-IV: Feedback and locking.}
The BS sends $(k^\star,l^\star)$ through downlink control information, and the target uses its low-power control receiver only for mode switching. The BS then locks $\mathbf w_{k^\star}(\Theta_{\mathrm a})$, while the target locks $\boldsymbol\Phi_{l^\star}(\Theta_{\min}^{\mathrm{BM}})$ for subsequent sensing or backscatter communication.

\subsection{Reliability-Constrained Parent Problem}
\label{subsec:parent_problem}

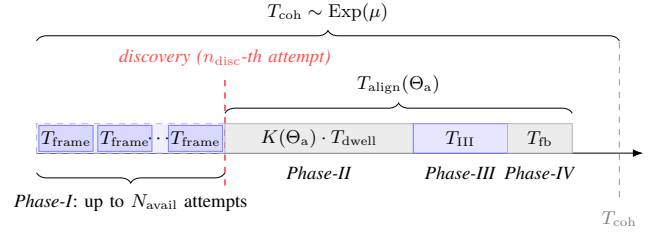
\begin{figure}[htbp]
\centering
\resizebox{\columnwidth}{!}{
\begin{tikzpicture}[>=Latex, font=\rmfamily\small]

    % ---- 尺寸参数 ----
    \def\blockH{0.50}
    \def\yMain{0}

    % Phase-I 最大窗口终点
    \def\phIstart{0}
    \def\phIend{3.20}

    % 发现事件位置
    \def\discX{3.20}

    % T_align 各段宽度
    \def\phIIW{3.20}
    \def\phIIIW{1.60}
    \def\phIVW{1.10}

    \pgfmathsetmacro{\phIIstart}{\discX}
    \pgfmathsetmacro{\phIIend}{\phIIstart+\phIIW}
    \pgfmathsetmacro{\phIIIstart}{\phIIend}
    \pgfmathsetmacro{\phIIIend}{\phIIIstart+\phIIIW}
    \pgfmathsetmacro{\phIVstart}{\phIIIend}
    \pgfmathsetmacro{\phIVend}{\phIVstart+\phIVW}

    % T_coh 终点
    \pgfmathsetmacro{\tcohEnd}{\phIVend+0.80}
    \pgfmathsetmacro{\axisEnd}{\tcohEnd+0.40}

    % ---- 主时间轴 ----
    \draw[->] (0, \yMain) -- (\axisEnd, \yMain);

    % ---- Phase-I 最大窗口底色 ----
    \draw[fill=blue!6, draw=blue!30, dashed]
        (\phIstart, \yMain) rectangle (\phIend, \blockH);

    % Phase-I 内部帧（调整间距避免省略号重叠）
    \draw[fill=blue!15, draw=blue!50]
        (0.04, 0.03) rectangle (0.96, \blockH-0.03);
    \node at (0.50, \blockH/2) {$T_{\mathrm{frame}}$};

    \draw[fill=blue!15, draw=blue!50]
        (1.04, 0.03) rectangle (1.96, \blockH-0.03);
    \node at (1.50, \blockH/2) {$T_{\mathrm{frame}}$};

    % 将省略号放在精确的中间位置
    \node at (2.10, \blockH/2) {$\cdots$};

    % 保持与前面框宽度相同的 0.92 宽度，给省略号留出空间
    \draw[fill=blue!15, draw=blue!50]
        (2.24, 0.03) rectangle (3.16, \blockH-0.03);
    \node at (2.70, \blockH/2) {$T_{\mathrm{frame}}$};

    % Phase-I 底部括号 (加大了 raise 和 below 参数)
    \draw[decorate, decoration={brace, amplitude=5pt, mirror, raise=18pt}]
        (\phIstart, \yMain+0.3) -- (\phIend, \yMain+0.3)
        node[midway, below=25pt]
       {\textit{Phase-I}:\ up to $N_{\mathrm{avail}}$ attempts};
        %{\textit{Phase-I}};

    % ---- 发现事件竖线 ----
    \draw[dashed, thick, color=red!70]
        (\discX, -0.55) -- (\discX, \blockH+0.85);
    \node[color=red!70, above, font=\rmfamily\small\itshape]
        at (\discX, \blockH+0.85)
        {discovery ($n_{\mathrm{disc}}$-th attempt)};

    % ---- Phase-II 块 ----
    \draw[fill=gray!15, draw=gray!60]
        (\phIIstart, \yMain) rectangle (\phIIend, \blockH);
    \node at (\phIIstart+\phIIW/2, \blockH/2)
        {$K(\Theta_{\mathrm{a}})\cdot T_{\mathrm{dwell}}$};
    \node[below=3pt] at (\phIIstart+\phIIW/2, \yMain)
        {\textit{Phase-II}};

    % ---- Phase-III 块 ----
    \draw[fill=blue!10, draw=blue!50]
        (\phIIIstart, \yMain) rectangle (\phIIIend, \blockH);
    \node at (\phIIIstart+\phIIIW/2, \blockH/2) {$T_{\mathrm{III}}$};
    \node[below=3pt] at (\phIIIstart+\phIIIW/2, \yMain)
        {\textit{Phase-III}};

    % ---- Phase-IV 块 ----
    \draw[fill=gray!15, draw=gray!60]
        (\phIVstart, \yMain) rectangle (\phIVend, \blockH);
    \node at (\phIVstart+\phIVW/2, \blockH/2) {$T_{\mathrm{fb}}$};
    \node[below=3pt] at (\phIVstart+\phIVW/2, \yMain)
        {\textit{Phase-IV}};

    % ---- T_align 括号（上方第一层）----
    \draw[decorate, decoration={brace, amplitude=5pt, raise=5pt}]
        (\phIIstart, \blockH) -- (\phIVend, \blockH)
        node[midway, above=11pt]
        {$T_{\mathrm{align}}(\Theta_{\mathrm{a}})$};

    % ---- T_coh 括号（上方第二层，大幅抬高避开红色文字）----
    \draw[decorate, decoration={brace, amplitude=5pt, raise=40pt}]
        (\phIstart, \blockH) -- (\tcohEnd, \blockH)
        node[midway, above=46pt]
        {$T_{\mathrm{coh}}\sim\mathrm{Exp}(\mu)$};

    % T_coh 终点虚线 (延长虚线以连接顶层大括号与底部基线)
    \draw[dashed, black!40]
        (\tcohEnd, -0.75) -- (\tcohEnd, \blockH+1.40);
    \node[black!50, below=3pt] at (\tcohEnd, -0.75) {$T_{\mathrm{coh}}$};

\end{tikzpicture}
}
\caption{Timeline of the proposed event-triggered discovery-and-locking protocol within one coherence window.}
\label{fig:timeline}
\end{figure}

We now formulate the parent design problem for establishing a high-quality locked link within a finite channel-coherence window. As illustrated in Fig.~\ref{fig:timeline}, the coherence time must accommodate downlink-assisted discovery, BS transmit sweeping, Backscatter-MIMO reflection sweeping, and feedback locking. Once the tagged echo is detected in Phase-I, the post-trigger alignment duration is
\begin{equation}\small
T_{\mathrm{align}}(\Theta_{\mathrm a})=K(\Theta_{\mathrm a})T_{\mathrm{dwell}}+T_{\mathrm{III}}+T_{\mathrm{fb}}.
\label{eq:T_align_protocol}
\end{equation}
If the useful discovery trigger occurs at the $n_{\mathrm{disc}}$-th downlink frame, the protocol can succeed only when
\begin{equation}\small
\mathcal{E}_{\mathrm{coh}}(\Theta_{\mathrm a})\triangleq\left\{n_{\mathrm{disc}}T_{\mathrm{frame}}+T_{\mathrm{align}}(\Theta_{\mathrm a})\leq T_{\mathrm{coh}}\right\}.
\label{eq:coh_completion_event}
\end{equation}
Equivalently, for a given realization of $T_{\mathrm{coh}}$, the number of useful Phase-I discovery opportunities is
\begin{equation}\small
N_{\mathrm{avail}}(\Theta_{\mathrm a},T_{\mathrm{coh}})=\left\lfloor\frac{[T_{\mathrm{coh}}-T_{\mathrm{align}}(\Theta_{\mathrm a})]^+}{T_{\mathrm{frame}}}\right\rfloor,
\label{eq:N_avail_protocol}
\end{equation}
where $[x]^+=\max\{x,0\}$.

% -------------------------------------------------------------------------

Using the locked-link SNR and beam-domain angular ambiguity characterized in Section~\ref{subsec:beamwidth_echo_model}, the parent problem is formulated over the beamwidths, waveform, and codebooks. Let $\mathcal{Z}\triangleq\{\Theta_{\mathrm d},\Theta_{\mathrm a},\boldsymbol{\alpha},\mathcal{W}(\Theta_{\mathrm a}),\mathcal{F}(\Theta_{\mathrm d}),\mathcal{F}(\Theta_{\min}^{\mathrm{BM}})\}$, where $\mathcal{F}(\Theta_{\mathrm d})$ is used in Phases~I--II and $\mathcal{F}(\Theta_{\min}^{\mathrm{BM}})$ is used in Phase~III. The parent problem is written as
\begin{equation}\small
\begin{aligned}
\text{(P1)}\quad \max_{\mathcal{Z}}\quad & \rho_{\mathrm{lock}}(\Theta_{\mathrm a},\Theta_{\min}^{\mathrm{BM}})\\
\mathrm{s.t.}\quad & \Pr\!\left\{\mathcal{D}_{\Theta_{\mathrm d}}\cap\mathcal{E}_{\mathrm{coh}}(\Theta_{\mathrm a})\cap\{k^\star=k_0\}\right\}\geq P_{\mathrm{req}},\\
& P_{\mathrm{FA}}(\Theta_{\mathrm d},\boldsymbol{\alpha})\leq P_{\mathrm{FA}}^{\mathrm{sys}},\\
& \Theta_{\mathrm d}\in[\Theta_{\min}^{\mathrm{BM}},\Omega],\quad \Theta_{\mathrm a}\in[\Theta_{\min}^{\mathrm{BS}},\Omega],\\
& \mathcal{W}(\Theta_{\mathrm a})\in\mathcal{C}_{\mathrm{BS}},\quad \mathcal{F}(\Theta_{\mathrm d}),\mathcal{F}(\Theta_{\min}^{\mathrm{BM}})\in\mathcal{C}_{\mathrm{BM}},\\
& \boldsymbol{\alpha}\in\mathcal{A}_{\mathrm{tag}}.
\end{aligned}
\label{eq:P1_parent}
\end{equation}
Here, $P_{\mathrm{req}}$ is the required protocol success probability, $P_{\mathrm{FA}}^{\mathrm{sys}}$ is the system-level false-alarm constraint, $\mathcal{C}_{\mathrm{BS}}$ and $\mathcal{C}_{\mathrm{BM}}$ denote the feasible BS and Backscatter-MIMO codebook sets, and $\mathcal{A}_{\mathrm{tag}}$ denotes the feasible tag-waveform set. The first constraint explicitly includes the coherence-time completion event $\mathcal{E}_{\mathrm{coh}}(\Theta_{\mathrm a})$, and will be evaluated in Section~\ref{sec:performance} through the coherence-averaged end-to-end success probability $\bar{P}_{\mathrm{e2e}}(\Theta_{\mathrm d},\Theta_{\mathrm a})$.

\begin{remark}[Beamwidth-induced time competition]
\label{rem:time_competition}
The two beamwidths affect the finite-coherence protocol through different mechanisms. The BS alignment beamwidth $\Theta_{\mathrm a}$ determines the deterministic post-trigger reservation $T_{\mathrm{align}}(\Theta_{\mathrm a})$ through $K(\Theta_{\mathrm a})$; narrowing $\Theta_{\mathrm a}$ improves locked-link SNR and beam-domain angular ambiguity, but leaves fewer useful discovery opportunities before the channel changes. By contrast, the default Backscatter-MIMO beamwidth $\Theta_{\mathrm d}$ does not affect $T_{\mathrm{align}}(\Theta_{\mathrm a})$, but controls the stochastic discovery process and tagged-echo reliability. A wider $\Theta_{\mathrm d}$ enables earlier discovery through larger broad-in/out coverage, whereas a narrower $\Theta_{\mathrm d}$ strengthens the tagged echo and reduces NLoS false-peak risk. Hence, $\Theta_{\mathrm a}$ mainly governs post-trigger latency and locked-link quality, while $\Theta_{\mathrm d}$ mainly governs pre-trigger discovery reliability.
\end{remark}

Since $\rho_{\mathrm{lock}}(\Theta_{\mathrm a},\Theta_{\min}^{\mathrm{BM}})$ increases as $\Theta_{\mathrm a}$ decreases, and $\sigma_{\theta,\mathrm{lock}}^2(\Theta_{\mathrm a})$ decreases as $\Theta_{\mathrm a}$ decreases, P1 seeks the narrowest reliability-feasible BS sweeping beamwidth rather than the beamwidth that merely maximizes the probability of completing a coarse sweep.

\subsection{Protocol-Guided Problem Decomposition}
\label{subsec:problem_decomposition}

The parent problem in \eqref{eq:P1_parent} couples the offline waveform design, the dual-end codebook construction, and the online protocol reliability. To make the problem tractable, we decompose it according to the protocol phases.

\noindent\textbf{P2: Dual-end adjustable-width codebook construction.}
The BS transmit codebook and the Backscatter-MIMO reflection codebook should realize controllable beamwidths over the prescribed angular region. For the BS side, the objective is to construct
\begin{equation}\small
\mathcal{W}(\Theta_{\mathrm a})=
\left\{
\mathbf{w}_{k}(\Theta_{\mathrm a})
\right\}_{k=1}^{K(\Theta_{\mathrm a})}
\in\mathcal{C}_{\mathrm{BS}},
\label{eq:P2a}
\end{equation}
where each codeword has an effective beamwidth close to $\Theta_{\mathrm a}$ and the union of all codeword supports covers the BS search region. For the Backscatter-MIMO side, the objective is to construct
\begin{equation}\small
\mathcal{F}(\Theta)=
\left\{
\boldsymbol{\Phi}_{l}(\Theta)
\right\}_{l=1}^{L(\Theta)}
\in\mathcal{C}_{\mathrm{BM}},
\label{eq:P2b}
\end{equation}
where $\Theta=\Theta_{\mathrm d}$ is used in Phases~I--II, and $\Theta=\Theta_{\min}^{\mathrm{BM}}$ is used in Phase~III. Therefore, P2 designs $\mathcal{W}(\Theta_{\mathrm a})$ and $\mathcal{F}(\Theta)$ to provide adjustable angular coverage, controlled array gain, and unambiguous retro-directional reflection. This problem will be solved in Section~\ref{sec:codebook_waveform_design}.

\noindent\textbf{P3: Detection-oriented remodulation waveform design.}
The remodulation waveform should maximize the probability of detecting the Backscatter-MIMO tagged echo under a prescribed default reflection beamwidth $\Theta_{\mathrm d}$. Based on the asynchronous remodulation model in Section~\ref{subsec:async_model}, the feasible waveform set is defined as
\begin{equation}\small
\mathcal{A}_{\mathrm{tag}}
\triangleq
\left\{
\boldsymbol{\alpha}:
\mathbf{U}_{\mathrm{sc}}^{H}\boldsymbol{\alpha}=\mathbf{0},
\ \|\boldsymbol{\alpha}\|_2^2=1
\right\},
\label{eq:A_tag}
\end{equation}
where $\mathbf{U}_{\mathrm{sc}}$ spans the unmodulated downlink-clutter subspace. The waveform design problem is then formulated as
\begin{equation}\small
\text{(P3)}\quad
\boldsymbol{\alpha}^{\star}(\Theta_{\mathrm d})
=
\arg\max_{\boldsymbol{\alpha}\in\mathcal{A}_{\mathrm{tag}}}
P_D(\Theta_{\mathrm d},\boldsymbol{\alpha}).
\label{eq:P3_waveform}
\end{equation}
This formulation explicitly links the offline tag design to Phase-I discovery reliability. In Section~\ref{sec:codebook_waveform_design}, \eqref{eq:P3_waveform} is converted into a tractable fractional-delay energy maximization problem, because under a fixed CFAR constraint, $P_D(\Theta_{\mathrm d},\boldsymbol{\alpha})$ increases with the tagged-echo noncentrality, which is determined by the effective waveform energy $\|\tilde{\boldsymbol{\alpha}}(\Delta\tau)\|_2^2$.

With P2 and P3 solved offline, the remaining task is to evaluate the reliability constraint in \eqref{eq:P1_parent}. Section~\ref{sec:performance} derives the Phase-I discovery probability $P_D(\Theta_{\mathrm d},\boldsymbol{\alpha}^{\star})$, the Phase-II incident-beam outage probability $P_{\mathrm{out}}(\Theta_{\mathrm a},\Theta_{\mathrm d})$, and the resulting end-to-end protocol success probability $\bar{P}_{\mathrm{e2e}}$. These results characterize the feasible beamwidth region of P1 and identify the narrowest reliable sweeping beamwidth.

\section{Offline Design of Codebooks and Remodulation Waveform}
\label{sec:codebook_waveform_design}

This section develops the offline components required by the proposed protocol. Specifically, P2 is solved by constructing adjustable-width BS transmit beams and Backscatter-MIMO retro-directional reflection patterns, while P3 is solved by designing a clutter-aware and asynchrony-robust remodulation waveform. These offline designs provide the codebooks and tag sequence used by the online discovery-and-locking protocol.

\subsection{Dual-End Adjustable-Width Codebook Design}
\label{subsec:p2_codebook_design}

The purpose of P2 is to construct dual-end codebooks with controllable angular coverage. At the BS side, the codebook should support adjustable transmit beamwidths so that the protocol can trade post-lock gain for sweeping latency. At the Backscatter-MIMO side, the reflection codebook should support both a default broad-in/out mode for discovery and tagged-echo enhancement, and a narrow retro-directional mode for passive reflection locking.

\noindent\textbf{BS adjustable-width transmit codebook.}
A conventional DFT codebook only provides the narrowest array beam and may require excessive sweeping time within a finite coherence window. To obtain a controllable BS beamwidth, we introduce a quadratic phase-spoiling factor into each transmit codeword. For a prescribed BS alignment beamwidth $\Theta_{\mathrm a}$, the $k$-th BS transmit codeword is constructed as
\begin{equation}\small
[\mathbf{w}_{k}(\Theta_{\mathrm a})]_{m}=\frac{1}{\sqrt{M_{\mathrm{ant}}}}e^{j\pi m\nu_{k}}e^{j\gamma_{\mathrm{req}}(\Theta_{\mathrm a})\left(m-\frac{M_{\mathrm{ant}}-1}{2}\right)^2},
\label{eq:bs_qps_codebook}
\end{equation}
where $m=0,\ldots,M_{\mathrm{ant}}-1$, and $\nu_k$ denotes the $k$-th beam-center direction in the directional-cosine domain. The required phase-spoiling factor is chosen as
\begin{equation}\small
\gamma_{\mathrm{req}}(\Theta_{\mathrm a})=\frac{\pi}{M_{\mathrm{ant}}^2}\sqrt{\left(\frac{\Theta_{\mathrm a}}{\Theta_{\min}^{\mathrm{BS}}}\right)^2-1},
\label{eq:gamma_req}
\end{equation}
where $\Theta_{\min}^{\mathrm{BS}}=2/M_{\mathrm{ant}}$ is the minimum BS beamwidth. For an angular search region with width $\Omega$, the number of BS sweeping beams is approximately $K(\Theta_{\mathrm a})=\lceil\Omega/\Theta_{\mathrm a}\rceil$.

For the quadratic phase-spoiled codeword in \eqref{eq:bs_qps_codebook}, the phase-spoiling factor $\gamma$ broadens the spatial-frequency support of the array response. The resulting effective beamwidth and peak array gain can be approximated as
\begin{equation}\small
\Theta(\gamma)\approx\Theta_{\min}^{\mathrm{BS}}\sqrt{1+\left(\frac{M_{\mathrm{ant}}^2\gamma}{\pi}\right)^2},\quad G_{\mathrm{array}}^{\mathrm{BS}}(\gamma)\approx\frac{2}{\Theta(\gamma)}.
\label{eq:bs_beamwidth_gain}
\end{equation}
This relation shows that the phase-spoiling factor directly controls the beamwidth--gain tradeoff. When $\gamma=0$, the codeword reduces to a DFT-like narrow beam with beamwidth $\Theta_{\min}^{\mathrm{BS}}$. Increasing $\gamma$ broadens the mainlobe and reduces the peak gain approximately according to spatial-energy conservation. By selecting $\gamma=\gamma_{\mathrm{req}}(\Theta_{\mathrm a})$ in \eqref{eq:gamma_req}, the BS codebook realizes the prescribed alignment beamwidth $\Theta_{\mathrm a}$. Hence, $\Theta_{\mathrm a}$ becomes a protocol-level control variable: a smaller $\Theta_{\mathrm a}$ improves the post-lock BS-side gain and angular resolution, whereas a larger $\Theta_{\mathrm a}$ reduces the number of BS sweeping beams.

% \begin{figure}[htbp]
%     \centering
%     \includegraphics[width=0.78\linewidth]{figs/bs_codebooks.pdf}
%     \caption{BS transmit beam patterns generated by the quadratic phase-spoiled codebook.}
%     \label{fig:bs_codebooks}
% \end{figure}

\noindent\textbf{Backscatter-MIMO retro-directional reflection codebook.}
The Backscatter-MIMO codebook has a different role from the BS transmit codebook. It should not only realize adjustable reflection beamwidths, but also return the incident energy toward the BS through a retro-directional phase profile. For an incident direction $\theta$, the round-trip spatial response across the Backscatter-MIMO array is
\begin{equation}\small
[\mathbf{b}(\theta)]_{n}=e^{-j\frac{4\pi}{\lambda}d_{\mathrm{BM}}n\sin\theta},\quad n=0,\ldots,N-1.
\label{eq:round_trip_response}
\end{equation}
The factor $4\pi/\lambda$ appears because the phase is accumulated over the incident and reflected paths. We set $d_{\mathrm{BM}}=\lambda/4$ so that the round-trip spatial frequency is unambiguous over the directional-cosine interval.

For a prescribed reflection beamwidth $\Theta$, the $l$-th Backscatter-MIMO reflection vector is constructed as
\begin{equation}\small
[\mathbf{u}_{l}(\Theta)]_{n}=e^{j\frac{4\pi}{\lambda}d_{\mathrm{BM}}n\nu_{l}}e^{j\xi_{\mathrm{req}}(\Theta)\left(n-\frac{N-1}{2}\right)^2},
\label{eq:bm_retro_codebook}
\end{equation}
where $\nu_l$ denotes the $l$-th reflection-codeword center in the directional-cosine domain. The corresponding reflection matrix is
$\boldsymbol{\Phi}_{l}(\Theta)=\mathrm{diag}\{\mathbf{u}_{l}(\Theta)\}$.
The required Backscatter-MIMO phase-spoiling factor is
\begin{equation}\small
\xi_{\mathrm{req}}(\Theta)=\frac{\pi}{N^2}\sqrt{\left(\frac{\Theta}{\Theta_{\min}^{\mathrm{BM}}}\right)^2-1},
\label{eq:xi_req}
\end{equation}
where $\Theta_{\min}^{\mathrm{BM}}=2/N$ is the narrowest retro-directional reflection beamwidth. The number of reflection configurations is approximately $L(\Theta)=\lceil\Omega/\Theta\rceil$.

The Backscatter-MIMO uses this codebook in two modes. In Phases~I--II, it adopts the default broad-in/out mode with $\Theta=\Theta_{\mathrm d}$, which enlarges the angular region over which the target can be discovered and provides tagged-echo enhancement during BS sweeping. In Phase~III, it switches to the narrowest retro-directional sweeping mode with $\Theta=\Theta_{\min}^{\mathrm{BM}}$, which concentrates the reflected energy and refines the passive reflection direction after the BS incident beam has been selected.

\begin{remark}[Role of retro-directional reflection]
\label{rem:retro_direction}
The retro-directional codebook differs from a conventional reflection codebook because its phase profile is designed to compensate the round-trip spatial response in \eqref{eq:round_trip_response}. For the correct reflection configuration, the tagged echo is coherently returned toward the BS. The quarter-wavelength spacing suppresses round-trip grating-lobe ambiguity, while the quadratic phase-spoiling factor controls the angular width of the retro-directional high-gain region.
\end{remark}

\begin{figure}[htbp]
    \centering
    \includegraphics[width=0.9\linewidth]{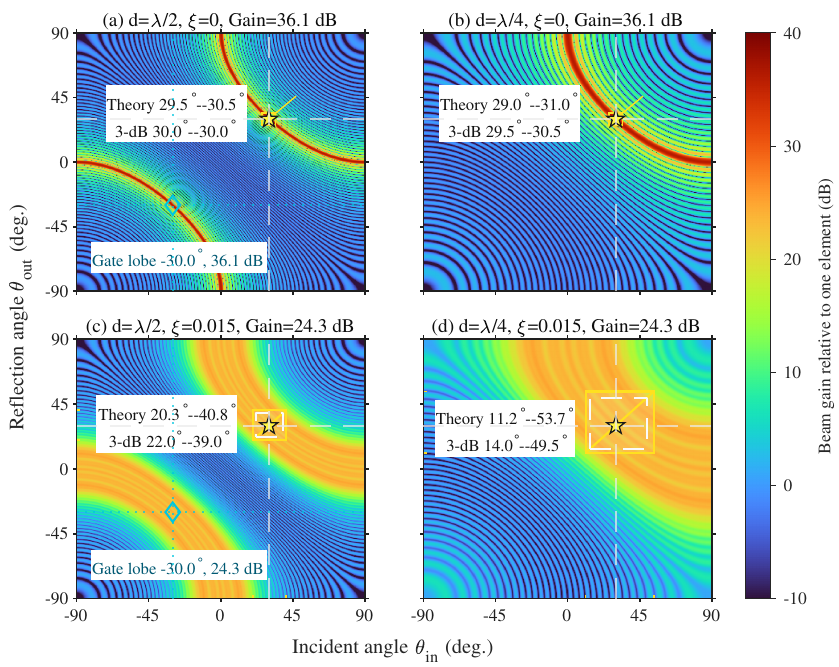}
    \caption{Retro-directional reflection patterns generated by the Backscatter-MIMO codebook. The solid boxes denote the theory-predicted retro-directional ranges, while the dashed boxes denote the numerically extracted high-gain ranges.}
    \label{fig:retro_direction_patterns}
\end{figure}

Fig.~\ref{fig:retro_direction_patterns} illustrates how the retro-directional reflection width changes with the phase-spoiling factor and element spacing. With $\xi=0$, the reflection pattern is narrow and provides the highest retro-directional gain, which is suitable for Phase-III direction locking but too selective for default discovery. Increasing $\xi$ broadens the retro-directional high-gain region, thereby improving angular coverage at the cost of reduced peak reflection gain; this is the operating principle behind the default broad-in/out mode with $\Theta_{\mathrm d}$. The comparison between $d_{\mathrm{BM}}=\lambda/2$ and $d_{\mathrm{BM}}=\lambda/4$ further shows that quarter-wavelength spacing removes the round-trip grating-lobe branch and yields a cleaner one-to-one retro-directional mapping.
Combining the BS and Backscatter-MIMO designs, P2 is solved by
$\mathcal{W}(\Theta_{\mathrm a})=\{\mathbf{w}_{k}(\Theta_{\mathrm a})\}_{k=1}^{K(\Theta_{\mathrm a})}$, $\mathcal{F}(\Theta)=\{\boldsymbol{\Phi}_{l}(\Theta)\}_{l=1}^{L(\Theta)}$,
where $\Theta=\Theta_{\mathrm d}$ is used in Phases~I--II and $\Theta=\Theta_{\min}^{\mathrm{BM}}$ is used in Phase~III.

\subsection{Detection-Oriented Remodulation Waveform Design}
\label{subsec:p3_waveform_design}

We now solve P3 by designing the remodulation waveform for downlink-assisted target discovery and tagged-echo extraction. For a prescribed default Backscatter-MIMO beamwidth $\Theta_{\mathrm d}$ and a fixed CFAR constraint, the discovery probability increases with the effective noncentrality of the matched-filter output. The waveform should therefore preserve the tagged-echo energy under fractional asynchronous sampling while suppressing the unmodulated downlink-clutter component.

From the cyclic asynchronous remodulation model, the effective sampled tag sequence can be written as
\begin{equation}\small
\tilde{\boldsymbol{\alpha}}_{k}(\Delta\tau)=\boldsymbol{\Pi}^{k}\mathbf{D}(\Delta\tau)\boldsymbol{\alpha},
\label{eq:alpha_tilde_matrix}
\end{equation}
where $\boldsymbol{\Pi}$ is the cyclic-delay matrix, and
\begin{equation}\small
\mathbf{D}(\Delta\tau)=\sum_{i=-\infty}^{+\infty}g_{\mathrm{bsm}}\!\left((i-\Delta\tau)T_s\right)\boldsymbol{\Pi}^{i}
\label{eq:D_delta}
\end{equation}
is the fractional-delay-induced cyclic ISI matrix. Under the adjacent-symbol approximation,
\begin{equation}\small
\mathbf{D}(\Delta\tau)\approx g_{\mathrm{bsm}}(-\Delta\tau T_s)\mathbf{I}+g_{\mathrm{bsm}}\!\left((1-\Delta\tau)T_s\right)\boldsymbol{\Pi}.
\label{eq:D_delta_approx}
\end{equation}
Since $\boldsymbol{\Pi}^{k}$ is unitary, the integer delay only cyclically shifts the tag sequence and does not change its energy. Thus, the waveform-dependent effective tag energy is
\begin{equation}\small
\left\|\tilde{\boldsymbol{\alpha}}_{k}(\Delta\tau)\right\|_2^2=\boldsymbol{\alpha}^{H}\mathbf{M}(\Delta\tau)\boldsymbol{\alpha},\quad \mathbf{M}(\Delta\tau)\triangleq\mathbf{D}^{H}(\Delta\tau)\mathbf{D}(\Delta\tau).
\label{eq:M_delta_def}
\end{equation}

Let $\bar{\mathbf{M}}\triangleq\mathbb{E}_{\Delta\tau}[\mathbf{M}(\Delta\tau)]$, where the expectation is taken over the fractional timing offset. To remove the dominant unmodulated clutter component, the tag waveform is constrained to lie in the orthogonal complement of the clutter subspace. With $\mathbf{U}_{\mathrm{sc}}$ denoting an orthonormal basis of the unmodulated downlink-clutter subspace, P3 is converted into the projected average-energy maximization problem
\begin{equation}\small
\begin{aligned}
\boldsymbol{\alpha}^{\star}=\arg\max_{\boldsymbol{\alpha}}\quad & \boldsymbol{\alpha}^{H}\bar{\mathbf{M}}\boldsymbol{\alpha}\\
\mathrm{s.t.}\quad & \mathbf{U}_{\mathrm{sc}}^{H}\boldsymbol{\alpha}=\mathbf{0},\quad \|\boldsymbol{\alpha}\|_2^2=1.
\end{aligned}
\label{eq:P3_energy}
\end{equation}
The null-space constraint suppresses the unmodulated clutter leakage, while the objective preserves the average tag energy after fractional-delay distortion.

Let $\mathbf{P}_{\perp}=\mathbf{I}-\mathbf{U}_{\mathrm{sc}}\mathbf{U}_{\mathrm{sc}}^{H}$. Problem~\eqref{eq:P3_energy} is solved by the principal eigenvector of the projected asynchronous-energy matrix:
\begin{equation}\small
\boldsymbol{\alpha}^{\star}=\mathbf{v}_{\max}\!\left(\mathbf{P}_{\perp}\bar{\mathbf{M}}\mathbf{P}_{\perp}\right).
\label{eq:alpha_star_solution}
\end{equation}
The maximum average effective tag energy is
\begin{equation}\small
\left(\boldsymbol{\alpha}^{\star}\right)^{H}\bar{\mathbf{M}}\boldsymbol{\alpha}^{\star}=\lambda_{\max}\!\left(\mathbf{P}_{\perp}\bar{\mathbf{M}}\mathbf{P}_{\perp}\right).
\label{eq:alpha_energy_max}
\end{equation}
Therefore, implementing P3 only requires an eigenvalue decomposition of $\mathbf{P}_{\perp}\bar{\mathbf{M}}\mathbf{P}_{\perp}$.

\begin{figure}[htbp]
    \centering
    \includegraphics[width=0.8\linewidth]{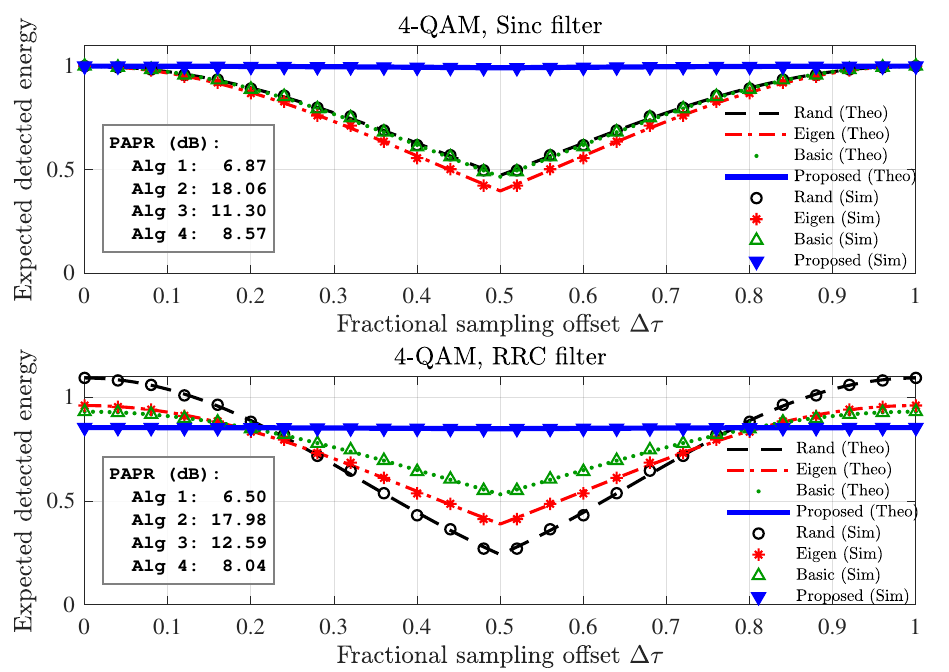}
    \caption{Detected tag energy versus fractional timing offset under different remodulation waveform designs.}
    \label{fig:remodulated_waveform_design}
\end{figure}

Fig.~\ref{fig:remodulated_waveform_design} verifies the fractional-delay robustness of the proposed remodulation waveform. The detected tag energy is evaluated over $\Delta\tau\in[0,1)$ and compared with three baselines: Rand, Basic, and Eigen. Rand denotes a random tag sequence, Basic denotes a null-space projected waveform without fractional-delay robustness, and Eigen denotes a direct eigenvector design without the same clutter-aware projected asynchronous-energy construction. Under both Sinc and RRC pulse-shaping filters, Rand and Basic suffer from severe energy degradation around fractional timing offsets, whereas the proposed waveform maintains a nearly flat energy profile. This confirms that $\bar{\mathbf{M}}$ captures the average fractional-delay distortion and that the projected eigenvector design preserves robust tag energy after clutter suppression.

\noindent With the codebooks in Section~\ref{subsec:p2_codebook_design} and the waveform in \eqref{eq:alpha_star_solution}, the offline components required by the proposed discovery-and-locking protocol are fully specified. Section~\ref{sec:performance} will use these designs to derive the discovery probability, alignment outage probability, and end-to-end protocol success probability.

\section{Protocol-Level Performance Analysis}
\label{sec:performance}
This section evaluates the reliability constraint in P1. We first derive the single-frame discovery probability $P_D(\Theta_{\mathrm d})$ and approximate the incident-beam alignment outage probability $P_{\mathrm{out}}(\Theta_{\mathrm a};\Theta_{\mathrm d})$. We then combine discovery, coherence-time completion, and incident-beam selection into the coherence-averaged success probability $\bar{P}_{\mathrm{e2e}}(\Theta_{\mathrm d},\Theta_{\mathrm a})$, which yields the reliability-constrained beamwidth selection rule for locked-link SNR and beam-domain angular ambiguity.

\subsection{Downlink-Assisted Discovery Probability}
\label{subsec:detection_prob}

In Phase-I, the BS reuses its ongoing downlink transmissions as the illuminating carrier. Since the downlink direction is determined by the primary communication schedule before target discovery, it is modeled as uniformly distributed over the angular sector $\Omega$. The Backscatter-MIMO target stays in its default broad-in/out reflection mode with beamwidth $\Theta_{\mathrm d}$, and the probability that the target is effectively illuminated by one downlink frame is $\Theta_{\mathrm d}/\Omega$.

Conditioned on this illumination event, the BS applies the tag-matched branch and suppresses the dominant unmodulated clutter subspace using the predesigned remodulation waveform. After clutter-aware projection, the detection model is approximated as
\begin{equation}\small
\mathbf{r}=\begin{cases}\tilde{\mathbf{n}}, & \mathcal{H}_0,\\ \sqrt{N_s^{\mathrm f}P_{\mathrm{tx}}}\mathbf{H}_{\mathrm{bsm}}(\Theta_{\mathrm d})\mathbf{w}\|\tilde{\boldsymbol{\alpha}}(\Delta\tau)\|+\tilde{\mathbf{n}}, & \mathcal{H}_1,\end{cases}
\label{eq:detection_hypothesis_projected}
\end{equation}
where $N_s^{\mathrm f}=\lfloor T_{\mathrm{frame}}f_s\rfloor$, $\tilde{\mathbf{n}}\sim\mathcal{CN}(\mathbf{0},\sigma^2\mathbf{I}_{M_{\mathrm{ant}}})$, and $\|\tilde{\boldsymbol{\alpha}}(\Delta\tau)\|$ denotes the effective tag-energy loss under fractional asynchronous sampling. Since the spatial signature is unknown before alignment, the BS uses the CFAR energy statistic $T=\|\mathbf{r}\|^2$.

\begin{lemma}[Single-frame discovery probability]
\label{lem:PD}
Consider the Phase-I detection model in \eqref{eq:detection_hypothesis_projected}. The CFAR threshold $T_{\mathrm{th}}$ satisfies
\begin{equation}\small
\frac{\Gamma(M_{\mathrm{ant}},T_{\mathrm{th}}/\sigma^2)}{\Gamma(M_{\mathrm{ant}})}=P_{\mathrm{FA}}^{\mathrm{sys}}.
\label{eq:CFAR_threshold}
\end{equation}
For an arbitrary feasible tag waveform $\boldsymbol{\alpha}$, the single-frame discovery probability is
\begin{equation}\small
P_D(\Theta_{\mathrm d},\boldsymbol{\alpha})=\frac{\Theta_{\mathrm d}}{\Omega}\mathbb{E}_{\Delta\tau}\!\left[Q_{M_{\mathrm{ant}}}\!\left(\sqrt{\rho(\Theta_{\mathrm d},\boldsymbol{\alpha},\Delta\tau)},\sqrt{\frac{2T_{\mathrm{th}}}{\sigma^2}}\right)\right],
\label{eq:PD_general}
\end{equation}
where $\rho(\Theta_{\mathrm d},\boldsymbol{\alpha},\Delta\tau)=\frac{4N_s^{\mathrm f}P_{\mathrm{tx}}|\beta_0|^4}{\sigma^2\Theta_{\mathrm d}}\|\tilde{\boldsymbol{\alpha}}(\Delta\tau)\|^2$.

For the optimized waveform $\boldsymbol{\alpha}^{\star}$, define $\lambda_{\alpha}\triangleq\lambda_{\max}(\mathbf{P}_{\perp}\bar{\mathbf{M}}\mathbf{P}_{\perp})$. Then, the optimized single-frame discovery probability is approximated as
\begin{equation}\small
P_D(\Theta_{\mathrm d})\approx\frac{\Theta_{\mathrm d}}{\Omega}Q_{M_{\mathrm{ant}}}\!\left(\sqrt{\bar{\rho}(\Theta_{\mathrm d})},\sqrt{\frac{2T_{\mathrm{th}}}{\sigma^2}}\right),
\label{eq:PD_opt}
\end{equation}
with
\begin{equation}\small
\bar{\rho}(\Theta_{\mathrm d})=\frac{4N_s^{\mathrm f}P_{\mathrm{tx}}|\beta_0|^4\lambda_{\alpha}}{\sigma^2\Theta_{\mathrm d}}.
\label{eq:rho_opt}
\end{equation}
\end{lemma}

The proof is given in Appendix~\ref{app:PD_derivation}. 
Under the no-target hypothesis, the matched-filter statistic contains only residual noise and follows a central Gamma distribution with shape parameter $M_{\mathrm{ant}}$ and scale parameter $\sigma^2$. Therefore, the ratio in \eqref{eq:CFAR_threshold} is the probability that noise alone exceeds $T_{\mathrm{th}}$. By setting this upper-tail probability to $P_{\mathrm{FA}}^{\mathrm{sys}}$, the detector satisfies the prescribed CFAR constraint independently of the target SNR and beamwidth choices. This also explains the detection tradeoff: lowering the allowed false-alarm probability increases the threshold and reduces the single-frame discovery probability. Lemma~\ref{lem:PD} shows that $\Theta_{\mathrm d}$ affects discovery through two opposite mechanisms. A wider default Backscatter-MIMO beam increases the illumination coverage $\Theta_{\mathrm d}/\Omega$, whereas a narrower one increases the tagged-echo non-centrality through the reflection gain. Thus, $\Theta_{\mathrm d}$ controls both discovery coverage and per-frame detection strength.

\subsection{Incident-Beam Alignment Outage Probability}
\label{subsec:outage_prob}

After Phase-I discovery, the BS switches to dedicated probing and sweeps the transmit codebook $\mathcal{W}(\Theta_{\mathrm a})$. Each BS codeword is used for a dwell time $T_{\mathrm{dwell}}$, which gives $N_s^{\mathrm{dwell}}=\lfloor T_{\mathrm{dwell}}f_s\rfloor$ samples per beam. During this phase, the Backscatter-MIMO target still uses its default reflection mode $\boldsymbol{\Phi}(\Theta_{\mathrm d})$ and imposes the optimized remodulation waveform $\boldsymbol{\alpha}^{\star}$. The BS selects the incident beam index according to $k^{\star}=\arg\max_k E_k$, where $E_k$ is the matched-filter tagged-echo energy associated with the $k$-th BS beam. An alignment outage occurs when $k^{\star}\neq k_0$.

\begin{lemma}[Incident-beam alignment outage probability]
\label{lem:Pout}
Consider Phase-II BS transmit sweeping with beamwidth $\Theta_{\mathrm a}$ and default Backscatter-MIMO reflection beamwidth $\Theta_{\mathrm d}$. The normalized desired LoS energy is
\begin{equation}\small
\gamma_{\mathrm{LoS}}(\Theta_{\mathrm a};\Theta_{\mathrm d})=\frac{4N_s^{\mathrm{dwell}}P_{\mathrm{tx}}\kappa|\beta_0|^4}{\sigma^2(\kappa+1)\Theta_{\mathrm a}\Theta_{\mathrm d}}.
\label{eq:gamma_LoS_explicit}
\end{equation}
The mean normalized NLoS-plus-noise energy of each misaligned beam is approximated as
\begin{equation}\small
\bar{\mu}_{\mathrm{NLoS}}(\Theta_{\mathrm a})=1+\frac{4N_s^{\mathrm{dwell}}P_{\mathrm{tx}}|\beta_0|^4}{\sigma^2(\kappa+1)\Theta_{\mathrm a}\Omega}.
\label{eq:mu_NLoS}
\end{equation}
Define the effective alignment SINR as $\frac{\gamma_{\mathrm{LoS}}(\Theta_{\mathrm a};\Theta_{\mathrm d})}{\bar{\mu}_{\mathrm{NLoS}}(\Theta_{\mathrm a})}$, we get
\begin{equation}\small
\eta(\Theta_{\mathrm a};\Theta_{\mathrm d})=\frac{\kappa|\beta_0|^4/\Theta_{\mathrm d}}{|\beta_0|^4/\Omega+\sigma^2(\kappa+1)\Theta_{\mathrm a}/(4N_s^{\mathrm{dwell}}P_{\mathrm{tx}})}.
\label{eq:eaSINR}
\end{equation}
Under the dominant-error and weakly correlated beam-bin approximations, the Phase-II incident-beam outage probability is
\begin{equation}\small
P_{\mathrm{out}}(\Theta_{\mathrm a};\Theta_{\mathrm d})\approx1-\left[1-\exp\!\left(-\eta(\Theta_{\mathrm a};\Theta_{\mathrm d})\right)\right]^{K(\Theta_{\mathrm a})-1},
\label{eq:Pout}
\end{equation}
where $K(\Theta_{\mathrm a})=\lceil\Omega/\Theta_{\mathrm a}\rceil$.
\end{lemma}

The proof is given in Appendix~\ref{app:Pout_derivation}. Lemma~\ref{lem:Pout} captures the false-peak competition in Phase-II. Narrowing $\Theta_{\mathrm a}$ increases per-beam spatial gain, but it also increases the number of competing beam hypotheses. Narrowing $\Theta_{\mathrm d}$ strengthens the desired tagged LoS echo after Phase-I discovery, while the average NLoS reflection level is governed by the full angular sector $\Omega$. Therefore, $P_{\mathrm{out}}$ acts as the reliability penalty imposed by NLoS-induced false-peak locking.

\subsection{End-to-End Protocol Success Probability}
\label{subsec:Pe2e_analysis}

The protocol succeeds if three events occur jointly: the tagged echo is discovered in Phase-I, the post-trigger alignment procedure is completed before the channel geometry changes, and the desired BS incident beam is selected in Phase-II. If discovery occurs at the $n$-th downlink frame, the trigger is useful only when $nT_{\mathrm{frame}}+T_{\mathrm{align}}(\Theta_{\mathrm a})\leq T_{\mathrm{coh}}$, where
\begin{equation}\small
T_{\mathrm{align}}(\Theta_{\mathrm a})=K(\Theta_{\mathrm a})T_{\mathrm{dwell}}+T_{\mathrm{III}}+T_{\mathrm{fb}}.
\label{eq:T_align_performance}
\end{equation}
Here, $T_{\mathrm{III}}$ includes the Backscatter-MIMO reflection-pattern sweep in Phase-III, and $T_{\mathrm{fb}}$ denotes the feedback-locking overhead in Phase-IV.

\begin{proposition}[Coherence-averaged end-to-end success probability]
\label{prop:Pe2e}
Let $T_{\mathrm{coh}}\sim\mathrm{Exp}(\mu)$. The coherence-averaged end-to-end protocol success probability is
\begin{equation}\small
\bar{P}_{\mathrm{e2e}}(\Theta_{\mathrm d},\Theta_{\mathrm a})=P_{\mathrm{trig}}(\Theta_{\mathrm d},\Theta_{\mathrm a})P_{\mathrm{align}}(\Theta_{\mathrm a};\Theta_{\mathrm d}),
\label{eq:Pe2e_factorized}
\end{equation}
where
\begin{equation}\small
P_{\mathrm{trig}}(\Theta_{\mathrm d},\Theta_{\mathrm a})=\frac{P_D(\Theta_{\mathrm d})e^{-\mu[T_{\mathrm{align}}(\Theta_{\mathrm a})+T_{\mathrm{frame}}]}}{1-[1-P_D(\Theta_{\mathrm d})]e^{-\mu T_{\mathrm{frame}}}},
\label{eq:Ptrig_closed}
\end{equation}
and
\begin{equation}\small
P_{\mathrm{align}}(\Theta_{\mathrm a};\Theta_{\mathrm d})=1-P_{\mathrm{out}}(\Theta_{\mathrm a};\Theta_{\mathrm d}).
\label{eq:Palign_closed}
\end{equation}
\end{proposition}

The proof is given in Appendix~\ref{app:Pe2e_proof}. The term $P_{\mathrm{trig}}$ is the probability that at least one useful discovery trigger occurs before the residual coherence time becomes insufficient for post-trigger alignment, and $P_{\mathrm{align}}$ is the probability that Phase-II selects the desired incident beam. Phase-III is not assigned a separate outage term because it is performed after the BS incident beam has been selected and the target is illuminated by a directional probing beam. Its time cost is included in $T_{\mathrm{align}}(\Theta_{\mathrm a})$, while the dominant pre-lock reliability loss is captured by $P_{\mathrm{out}}(\Theta_{\mathrm a};\Theta_{\mathrm d})$.

\subsection{Reliability-Constrained Beamwidth Design Law}
\label{subsec:beamwidth_design_law}

We now use \eqref{eq:Pe2e_factorized} to close the parent problem P1. Since $\rho_{\mathrm{lock}}(\Theta_{\mathrm a},\Theta_{\min}^{\mathrm{BM}})$ increases and $\sigma_{\theta,\mathrm{lock}}^2(\Theta_{\mathrm a})$ decreases as $\Theta_{\mathrm a}$ decreases, P1 is equivalent to finding the narrowest BS sweeping beamwidth whose reliability envelope satisfies the required end-to-end success probability.

Define the feasible default-beamwidth set as
\begin{equation}\small
\mathcal{D}_{\mathrm{FA}}\triangleq\left\{\Theta_{\mathrm d}\in[\Theta_{\min}^{\mathrm{BM}},\Omega]:P_{\mathrm{FA}}(\Theta_{\mathrm d},\boldsymbol{\alpha}^{\star})\leq P_{\mathrm{FA}}^{\mathrm{sys}}\right\}.
\label{eq:D_FA_def}
\end{equation}
For each $\Theta_{\mathrm a}$, define the reliability envelope
\begin{equation}\small
\Psi(\Theta_{\mathrm a})\triangleq\max_{\Theta_{\mathrm d}\in\mathcal{D}_{\mathrm{FA}}}\bar{P}_{\mathrm{e2e}}(\Theta_{\mathrm d},\Theta_{\mathrm a}).
\label{eq:Pe2e_envelope}
\end{equation}
Then, the exact reliability-constrained solution is characterized by
\begin{equation}\small
\Theta_{\mathrm a}^{\star}=\inf\left\{\Theta_{\mathrm a}\in[\Theta_{\min}^{\mathrm{BS}},\Omega]:\Psi(\Theta_{\mathrm a})\geq P_{\mathrm{req}}\right\},
\label{eq:theta_a_exact}
\end{equation}
and the corresponding default Backscatter-MIMO beamwidth is selected as
\begin{equation}\small
\Theta_{\mathrm d}^{\star}\in\arg\max_{\Theta_{\mathrm d}\in\mathcal{D}_{\mathrm{FA}}}\bar{P}_{\mathrm{e2e}}(\Theta_{\mathrm d},\Theta_{\mathrm a}^{\star}).
\label{eq:theta_d_exact}
\end{equation}
 \eqref{eq:theta_a_exact}--\eqref{eq:theta_d_exact} give the exact reliability-envelope characterization. Since $\bar{P}_{\mathrm{e2e}}$ contains the Marcum-$Q$ discovery term, the exponential coherence-time penalty, and the extreme-value outage term, a simple elementary closed-form optimizer is generally unavailable.

To obtain analytical design insight, consider the high-discovery-SNR regime where the Marcum-$Q$ term in \eqref{eq:PD_opt} is close to one within the useful discovery region. Then
\begin{equation}\small
P_D(\Theta_{\mathrm d})\approx\frac{\Theta_{\mathrm d}}{\Omega}.
\label{eq:PD_high_snr_approx}
\end{equation}
Also approximate $K(\Theta_{\mathrm a})\approx\Omega/\Theta_{\mathrm a}$ and
\begin{equation}\small
T_{\mathrm{align}}(\Theta_{\mathrm a})\approx\frac{\Omega T_{\mathrm{dwell}}}{\Theta_{\mathrm a}}+T_{\mathrm{III}}+T_{\mathrm{fb}}.
\label{eq:Talign_continuous}
\end{equation}
Let $p_{\mathrm{trig}}$ and $p_{\mathrm{align}}$ be two reliability budgets satisfying $p_{\mathrm{trig}}p_{\mathrm{align}}\geq P_{\mathrm{req}}$. Define $A(\Theta_{\mathrm a})\triangleq e^{-\mu[T_{\mathrm{align}}(\Theta_{\mathrm a})+T_{\mathrm{frame}}]}$ and $B\triangleq e^{-\mu T_{\mathrm{frame}}}$. The triggering constraint $P_{\mathrm{trig}}\geq p_{\mathrm{trig}}$ gives the lower bound on $\Theta_{\mathrm{d}}$
\begin{equation}\small
\Theta_{\mathrm d}^{\mathrm L}(\Theta_{\mathrm a})\triangleq\Omega\frac{p_{\mathrm{trig}}(1-B)}{A(\Theta_{\mathrm a})-p_{\mathrm{trig}}B},
\label{eq:theta_d_lower}
\end{equation}
provided that $A(\Theta_{\mathrm a})>p_{\mathrm{trig}}B$. This bound is imposed by timely discovery and therefore favors a sufficiently wide default Backscatter-MIMO mode.

Similarly, the alignment constraint $P_{\mathrm{align}}\geq p_{\mathrm{align}}$ is equivalent to
\begin{equation}\small
\eta(\Theta_{\mathrm a};\Theta_{\mathrm d})\geq\eta_{\mathrm{req}}(\Theta_{\mathrm a},p_{\mathrm{align}})\triangleq-\ln\!\left[1-p_{\mathrm{align}}^{1/(\Omega/\Theta_{\mathrm a}-1)}\right].
\label{eq:eta_req}
\end{equation}
Substituting \eqref{eq:eaSINR} gives the upper bound on $\Theta_{\mathrm{d}}$
\begin{equation}\small
\Theta_{\mathrm{d}}^{\mathrm{U}}(\Theta_{\mathrm{a}}) \triangleq
\frac{\kappa|\beta_0|^4}{
\eta_{\mathrm{req}}(\Theta_{\mathrm{a}},p_{\mathrm{align}})
\left(\dfrac{|\beta_0|^4}{\Omega}+\dfrac{\sigma^2(\kappa+1)\Theta_{\mathrm{a}}}{4N_s^{\mathrm{dwell}}P_{\mathrm{tx}}}\right)}.
\label{eq:theta_d_upper}
\end{equation}
This upper bound is imposed by incident-beam reliability and therefore favors a sufficiently narrow default Backscatter-MIMO mode. A feasible default beamwidth exists only when
\begin{equation}\small
\max\{\Theta_{\min}^{\mathrm{BM}},\Theta_{\mathrm d}^{\mathrm L}(\Theta_{\mathrm a})\}\leq\min\{\Omega,\Theta_{\mathrm d}^{\mathrm U}(\Theta_{\mathrm a})\}.
\label{eq:theta_d_feasibility}
\end{equation}
Consequently, an asymptotic design rule for P1 is
\begin{equation}\small
\Theta_{\mathrm a}^{\star}\approx\inf\left\{\Theta_{\mathrm a}\in[\Theta_{\min}^{\mathrm{BS}},\Omega]:\eqref{eq:theta_d_feasibility}\ \text{holds}\right\}.
\label{eq:theta_a_asymptotic}
\end{equation}
Once $\Theta_{\mathrm a}^{\star}$ is determined, a reliability-feasible default beamwidth can be chosen as
\begin{equation}\small
\Theta_{\mathrm d}^{\star}\approx\left[\Theta_{\mathrm d}^{\mathrm U}(\Theta_{\mathrm a}^{\star})\right]_{\Theta_{\mathrm d}^{\mathrm L}(\Theta_{\mathrm a}^{\star})}^{\Omega},
\label{eq:theta_d_asymptotic}
\end{equation}
where $[x]_a^b\triangleq\min\{\max\{x,a\},b\}$.

The above expressions summarize the protocol-level beamwidth principle. The lower bound $\Theta_{\mathrm d}^{\mathrm L}$ comes from timely discovery and favors a wider default Backscatter-MIMO reflection mode. The upper bound $\Theta_{\mathrm d}^{\mathrm U}$ comes from incident-beam reliability and favors a narrower default mode. Therefore, the optimal BS sweeping beamwidth $\Theta_{\mathrm a}^{\star}$ is the narrowest value for which these two requirements overlap, which maximizes the locked-link SNR and reduces beam-domain angular ambiguity while satisfying the end-to-end reliability constraint.

\section{Simulation Results}
\label{sec:simulation}
This section validates the proposed blind dual-end alignment protocol and the analysis in Section~\ref{sec:performance}. Unless otherwise specified, the BS has $M_{\mathrm{ant}}=32$ half-wavelength-spaced antennas, and the Backscatter-MIMO target has $N=32$ quarter-wavelength-spaced passive elements. The carrier frequency is $28$ GHz, the search sector is $\Omega=2.0$ rad, the target-related channel contains one LoS and $P=8$ NLoS paths, and the environment has $Q=8$ dominant static scatterers. The Rician factor $\kappa$ controls the LoS/NLoS ratio, with $1/\kappa$ indicating NLoS severity, and all probability results are averaged over independent channel, clutter, timing-offset, and noise realizations. The benchmark schemes are non-cooperative energy detection with fixed broad beams (B1), cooperative RSSI feedback with fixed narrow beams (B2), blind random scanning with random tags and reflection states (B3), and CS-based beam alignment using sparse angular recovery (B4). The proposed scheme selects $(\Theta_{\mathrm d}^{\star},\Theta_{\mathrm a}^{\star})$ according to Section~\ref{subsec:beamwidth_design_law}.

\subsection{Protocol Mechanisms and Beamwidth Impact}
\label{subsec:sim_protocol_mechanism}
This subsection first visualizes how the proposed protocol converts the Backscatter-MIMO structural correspondence into an executable alignment procedure, and then examines the distinct roles of the two protocol beamwidths.

\begin{figure}[htbp]
    \centering
    \includegraphics[width=0.8\linewidth]{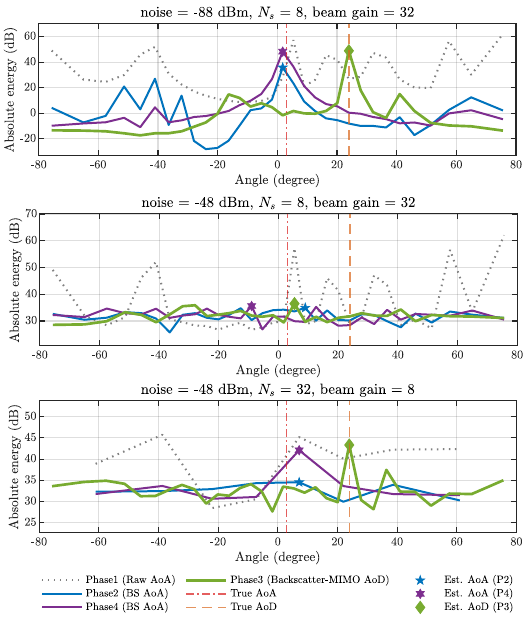}
    \caption{Protocol-level angular-spectrum evolution.}
    \label{fig:angular_spectrum_evolution}
\end{figure}
Fig.~\ref{fig:angular_spectrum_evolution} illustrates the angular-spectrum evolution of the proposed protocol under three representative regimes. The dotted gray curve denotes the raw Phase-I spectrum before tag extraction, while the blue, green, and purple curves correspond to the Phase-II BS search, the Phase-III Backscatter-MIMO reflection search, and the Phase-IV locked BS-side spectrum, respectively. The dashed vertical lines mark the true AoA and AoD, and the markers denote the estimates.
In the high-SNR narrow-beam case, the raw spectrum is dominated by static clutter and does not reveal the target direction. Reflection-modulated extraction converts it into a tagged-echo spectrum with a clear AoA peak, verifying waveform-domain clutter separation. The subsequent reflection sweep yields a sharp AoD peak, and the locked spectrum remains concentrated, showing that retro-directional passive beamforming provides reliable spatial focusing when the link budget is sufficient.
In the low-SNR narrow-beam case, reflection modulation still removes the dominant clutter, but the Phase-II and Phase-IV spectra fluctuate at comparable levels, the AoA estimate deviates, and the Phase-III search lacks a dominant true-AoD peak. Thus, clutter separation alone is insufficient under limited SNR and finite dwell time, where narrow exhaustive sweeping remains vulnerable to noise and NLoS false peaks.
The low-SNR broad-beam case restores alignment by reducing the search burden and increasing the effective accumulation time per beam. Although the peak gain is lower, the spectra become more stable and the estimated Backscatter-MIMO AoD approaches the true direction. Hence, Fig.~\ref{fig:angular_spectrum_evolution} confirms the two core mechanisms: reflection modulation makes the target echo distinguishable from clutter, while retro-directional and adjustable-width beam searches suppress NLoS false-peak locking.

 \begin{figure}[htbp]
    \centering
    \includegraphics[width=0.86\linewidth]{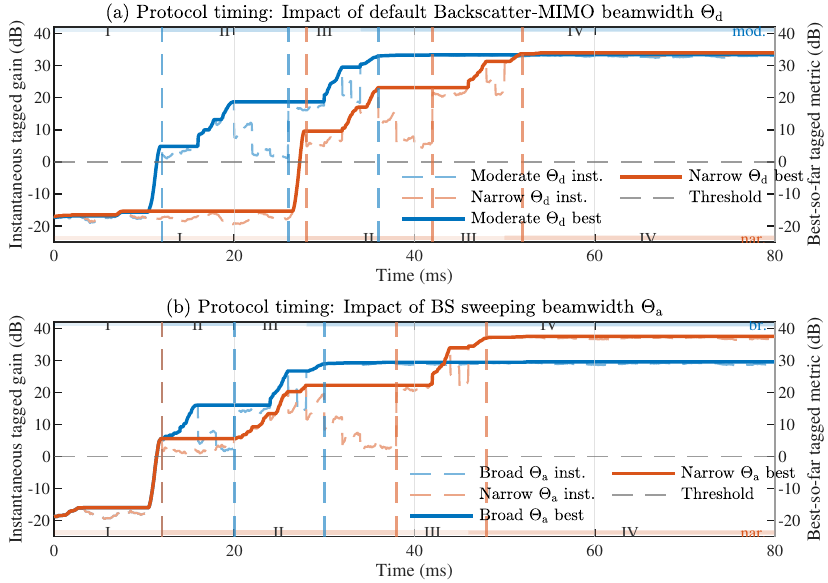}
    \caption{Protocol-level time evolution under different beamwidths.}
    \label{fig:protocol_time_evolution}
\end{figure}
Fig.~\ref{fig:protocol_time_evolution} examines the protocol-level time behavior. In Fig.~\ref{fig:protocol_time_evolution}(a), $\Theta_{\mathrm a}$ is fixed and two values of $\Theta_{\mathrm d}$ are compared. A wider default Backscatter-MIMO beam improves broad-in/out coverage and tends to trigger discovery earlier, while a narrower default beam provides stronger tagged echoes but may delay the first useful discovery opportunity. Once discovery is triggered, the post-trigger alignment duration is almost unchanged because $\Theta_{\mathrm a}$ is fixed.
In Fig.~\ref{fig:protocol_time_evolution}(b), $\Theta_{\mathrm d}$ is fixed and two values of $\Theta_{\mathrm a}$ are compared. The discovery behavior remains nearly unchanged, while the alignment duration and final locked gain change significantly. A broader $\Theta_{\mathrm a}$ shortens the sweep but yields lower post-lock quality, whereas a narrower $\Theta_{\mathrm a}$ increases the sweep time but improves the locked-link gain when the coherence-time constraint is satisfied. These results support the formulation in P1: $\Theta_{\mathrm d}$ mainly governs stochastic discovery reliability, while $\Theta_{\mathrm a}$ governs deterministic post-trigger latency and locked-link SNR and beam-domain angular ambiguity.

\subsection{Discovery and Alignment Reliability Metrics}
\label{subsec:sim_reliability_metrics}
This subsection validates the two elementary reliability metrics derived in Section~\ref{sec:performance}: the single-frame discovery probability and the incident-beam alignment outage probability. These two metrics isolate the discovery bottleneck caused by clutter and the alignment bottleneck caused by NLoS-induced false peaks.
\begin{figure}[htbp]
    \centering
    \includegraphics[width=0.8\linewidth]{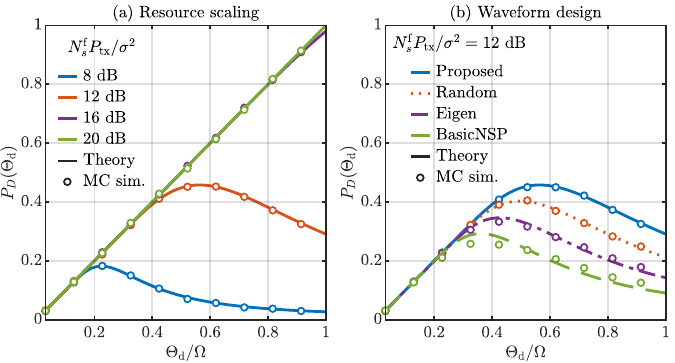}
    \caption{ 
    (a) Single-frame discovery probability under different sensing resources. 
    (b) Discovery probability of different remodulation waveform designs under the same sensing resource. }
    \label{fig:PD_validation}
\end{figure}
In Fig.~\ref{fig:PD_validation}(a), $P_D(\Theta_{\mathrm d})$ is plotted versus $\Theta_{\mathrm d}/\Omega$ under different $N_s^{\mathrm f}P_{\mathrm{tx}}/\sigma^2$, with analytical curves from \eqref{eq:PD_opt} and Monte Carlo markers. The close agreement verifies the non-central chi-square detection approximation. More importantly, $P_D(\Theta_{\mathrm d})$ is non-monotonic: a very narrow $\Theta_{\mathrm d}$ provides strong reflection gain but insufficient illumination coverage, whereas a very wide $\Theta_{\mathrm d}$ improves coverage but weakens the tagged-echo noncentrality. Hence, the discovery beamwidth is determined by the competition between coverage probability and reflected-echo strength.
Fig.~\ref{fig:PD_validation}(b) compares different remodulation waveform designs under the same sensing resource. The proposed P3 waveform achieves the highest discovery probability over the beamwidth sweep because it maximizes the fractional-delay averaged effective tag energy within the clutter-null subspace. In contrast, Random lacks both clutter-aware and timing-robust structure, while Eigen and BasicNSP suppress clutter but do not optimize the asynchronous effective-energy profile. This confirms that the proposed waveform design improves discovery reliability.

\begin{figure}[htbp]
    \centering
    \includegraphics[width=0.8\linewidth]{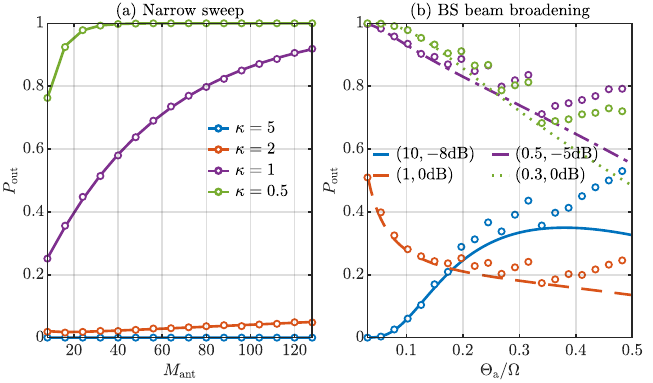}
    \caption{ 
    (a) Outage probability under narrow exhaustive BS sweeping as the BS array size increases. 
    (b) Regime-dependent effect of BS beam broadening.}
    \label{fig:Pout_validation}
\end{figure}
Fig.~\ref{fig:Pout_validation}(a) plots $P_{\mathrm{out}}(\Theta_{\min}^{\mathrm{BS}};\Theta_{\mathrm d})$ versus $M_{\mathrm{ant}}$ under different Rician factors $\kappa$. The Monte Carlo markers closely follow the analytical curves, confirming the dominant-error approximation in \eqref{eq:Pout}. As $\kappa$ decreases, the NLoS false peaks become more competitive, and the outage probability increases rapidly with the array size. This shows that exhaustive narrow-beam sweeping is not always improved by larger arrays: although narrower beams provide higher directional gain, they also create more competing beam hypotheses, leading to an array-scaling paradox in NLoS-dominated regimes.
Fig.~\ref{fig:Pout_validation}(b) further illustrates the regime-dependent effect of BS beam broadening. In Fig.~\ref{fig:Pout_validation}(b), each legend entry $(\kappa,\rho_{\mathrm{dw}})$ uses $\rho_{\mathrm{dw}}=10\log_{10}(N_s^{\mathrm{dwell}}P_{\mathrm{tx}}/\sigma^2)$ in dB.
The analytical curves use the continuous approximation $K=\Omega/\Theta_{\mathrm a}$, while the Monte Carlo markers use the integer codebook size $K=\lceil\Omega/\Theta_{\mathrm a}\rceil$, which explains the visible marker deviations caused by discrete beam-bin changes. In NLoS-dominated or resource-limited regimes, increasing $\Theta_{\mathrm a}$ reduces the number of competing false-peak bins and improves alignment reliability. In contrast, under LoS-favorable but finite-resource conditions, excessive broadening weakens the desired LoS metric and may increase outage. Therefore, BS beamwidth should be selected according to the reliability constraint rather than broadened unconditionally.

\subsection{End-to-End Feasibility and Locked-Link Performance}
\label{subsec:sim_e2e_benchmark}
This subsection evaluates the protocol-level consequence of the preceding reliability metrics. We first examine how discovery and alignment jointly determine the end-to-end reliability-feasible beamwidth region, and then compare the resulting reliability-gated locked-link performance with representative baselines.
\begin{figure}[htbp]
    \centering
    \includegraphics[width=0.8\linewidth]{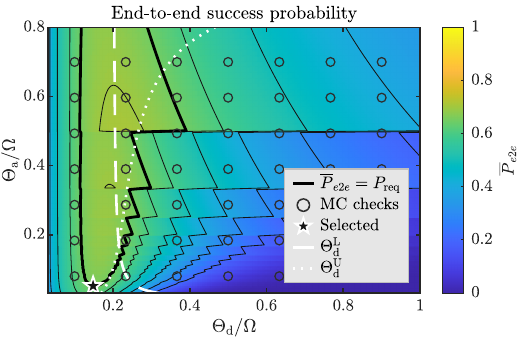}
    \caption{Coherence-averaged end-to-end success probability with reliability constraints over the default Backscatter-MIMO beamwidth $\Theta_{\mathrm d}$ and the BS sweeping beamwidth $\Theta_{\mathrm a}$.}
    \label{fig:e2e_heatmap}
\end{figure}
Fig.~\ref{fig:e2e_heatmap} validates Proposition~1 and illustrates the reliability-constrained beamwidth design law. The heatmap shows the coherence-averaged end-to-end success probability $\bar P_{\mathrm{e2e}}(\Theta_{\mathrm d},\Theta_{\mathrm a})$ over the default Backscatter-MIMO beamwidth and the BS sweeping beamwidth. The solid contour gives the exact reliability boundary $\bar P_{\mathrm{e2e}}=P_{\mathrm{req}}$, while the two white asymptotic bounds explain the feasible region: $\Theta_{\mathrm d}^{\mathrm L}$ is the discovery-induced lower bound, so feasible designs should lie to its right, whereas $\Theta_{\mathrm d}^{\mathrm U}$ is the alignment-induced upper bound, so overly wide default Backscatter-MIMO beams are excluded. The selected point is not the global maximizer of $\bar P_{\mathrm{e2e}}$; instead, it is the smallest feasible $\Theta_{\mathrm a}$ satisfying the reliability constraint, consistent with the fact that a narrower locked BS beam yields a higher final locked-link gain. This confirms that the proposed design law captures the key tradeoff among discovery latency, alignment outage, and coherence-time pressure.

\begin{figure}[htbp]
    \centering
    \includegraphics[width=0.85\linewidth]{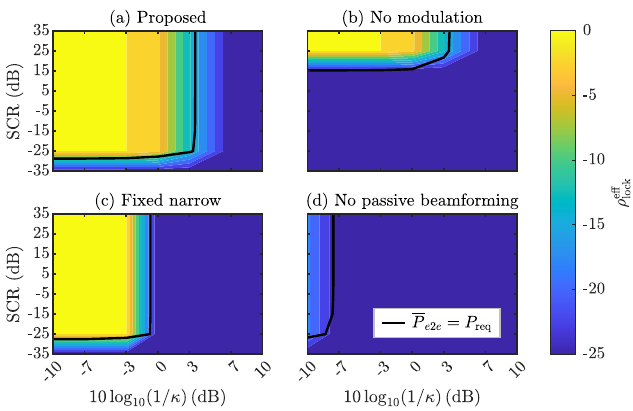}
    \caption{reliability-gated locked-link SNR under different signal-to-clutter ratios and NLoS severities. The horizontal axis is $10\log_{10}(1/\kappa)$, and the color denotes the reliability-gated locked-link SNR $\rho_{\mathrm{lock}}^{\mathrm{eff}}$ in dB. The black contour marks $\bar P_{\mathrm{e2e}}=P_{\mathrm{req}}$.}
    \label{fig:benchmark_lock_quality}
\end{figure}
Fig.~\ref{fig:benchmark_lock_quality} summarizes the reliability-gated post-lock quality under different SCRs and NLoS severities. The color represents the normalized reliability-gated locked-link SNR $\rho_{\mathrm{lock}}^{\mathrm{eff}}$ in dB, where the post-lock quality is counted only when the end-to-end reliability satisfies $\bar P_{\mathrm{e2e}}\geq P_{\mathrm{req}}$. The proposed design maintains the largest high-quality feasible region, since reflection modulation suppresses clutter in the low-SCR regime, while beamwidth selection mitigates NLoS false-peak locking without sacrificing the locked-link gain. Without reflection modulation, the feasible region shrinks significantly at low SCR, showing that clutter suppression is essential for reliable blind discovery. Fixed narrow sweeping preserves the highest potential lock quality, but becomes unreliable as $10\log_{10}(1/\kappa)$ increases because the exhaustive narrow-beam scan is more vulnerable to NLoS false peaks. In contrast, the no-passive-beamforming baseline remains weak even with reflection modulation, confirming that the Backscatter-MIMO retro-directional reflection codebook is necessary to obtain reliable LoS locking and high post-lock quality in cluttered multipath channels.

\section{Conclusion}
\label{sec:conclusion}
This paper investigated reliable standalone Backscatter-MIMO alignment in cluttered multipath ISAC environments. We showed that this problem cannot be addressed by conventional beam sweeping alone: the weak target-coupled echo must first be made distinguishable from strong environmental clutter, and dual-end beam locking must then be completed before the spatial channel becomes outdated. To this end, we developed a reflection-modulated acquisition and reliability-constrained alignment framework that jointly exploits tagged-echo separation and retro-directional passive beamforming.
The analysis and simulations showed that beamwidth is a protocol-level reliability variable, rather than only a spatial-resolution parameter. Narrow beams improve locked-link gain and angular resolution, but may increase sweeping latency and NLoS false-peak vulnerability; broad beams improve search robustness, but sacrifice locked-link quality. The resulting design principle is to choose the narrowest BS sweeping beamwidth that remains feasible under the end-to-end reliability constraint. More broadly, these results suggest a cooperative target-sensing paradigm in which surface-mounted Backscatter-MIMO arrays turn otherwise passive targets into programmable sensing participants, opening new opportunities for ISAC applications beyond conventional environmental sensing.

\appendices
\section{Derivation of Lemma~\ref{lem:PD}}
\label{app:PD_derivation}

Under $\mathcal{H}_0$, \eqref{eq:detection_hypothesis_projected} gives $\mathbf{r}=\tilde{\mathbf{n}}$, where $\tilde{\mathbf{n}}\sim\mathcal{CN}(\mathbf{0},\sigma^2\mathbf{I}_{M_{\mathrm{ant}}})$. Hence, $T=\|\mathbf{r}\|^2$ satisfies $T/\sigma^2\sim\mathrm{Gamma}(M_{\mathrm{ant}},1)$, and
\begin{equation}\small
P_{\mathrm{FA}}=\Pr\{T>T_{\mathrm{th}}|\mathcal{H}_0\}=\frac{\Gamma(M_{\mathrm{ant}},T_{\mathrm{th}}/\sigma^2)}{\Gamma(M_{\mathrm{ant}})}.
\label{eq:app_PFA}
\end{equation}
Setting $P_{\mathrm{FA}}=P_{\mathrm{FA}}^{\mathrm{sys}}$ gives \eqref{eq:CFAR_threshold}.

Under $\mathcal{H}_1$, conditioned on $\Delta\tau$, \eqref{eq:detection_hypothesis_projected} gives $\mathbf{r}=\mathbf{s}_{\mathrm d}+\tilde{\mathbf{n}}$, where
$\mathbf{s}_{\mathrm d}=\sqrt{N_s^{\mathrm f}P_{\mathrm{tx}}}\mathbf{H}_{\mathrm{bsm}}(\Theta_{\mathrm d})\mathbf{w}\|\tilde{\boldsymbol{\alpha}}(\Delta\tau)\|$.

Therefore, $2T/\sigma^2$ follows a non-central chi-square distribution with $2M_{\mathrm{ant}}$ degrees of freedom and non-centrality parameter $2\|\mathbf{s}_{\mathrm d}\|^2/\sigma^2$. Using the default Backscatter-MIMO reflection-gain approximation, this parameter is approximated as
$\frac{4N_s^{\mathrm f}P_{\mathrm{tx}}|\beta_0|^4}{\sigma^2\Theta_{\mathrm d}}\|\tilde{\boldsymbol{\alpha}}(\Delta\tau)\|^2=\rho(\Theta_{\mathrm d},\boldsymbol{\alpha},\Delta\tau)$.
Thus,
\begin{equation}\small
\Pr\{T>T_{\mathrm{th}}|\mathcal{H}_1,\Delta\tau\}=Q_{M_{\mathrm{ant}}}\!\left(\sqrt{\rho(\Theta_{\mathrm d},\boldsymbol{\alpha},\Delta\tau)},\sqrt{\frac{2T_{\mathrm{th}}}{\sigma^2}}\right).
\label{eq:app_PD_cond}
\end{equation}
Multiplying \eqref{eq:app_PD_cond} by the illumination probability $\Theta_{\mathrm d}/\Omega$ and averaging over $\Delta\tau$ gives \eqref{eq:PD_general}.

For the optimized waveform, $\|\tilde{\boldsymbol{\alpha}}(\Delta\tau)\|^2=\boldsymbol{\alpha}^H\mathbf{M}(\Delta\tau)\boldsymbol{\alpha}$ and $\bar{\mathbf{M}}=\mathbb{E}_{\Delta\tau}[\mathbf{M}(\Delta\tau)]$. Since $\boldsymbol{\alpha}^{\star}$ is the principal eigenvector of $\mathbf{P}_{\perp}\bar{\mathbf{M}}\mathbf{P}_{\perp}$ in the feasible subspace,
\begin{equation}\small
\mathbb{E}_{\Delta\tau}\!\left[\|\tilde{\boldsymbol{\alpha}}^{\star}(\Delta\tau)\|^2\right]=(\boldsymbol{\alpha}^{\star})^H\bar{\mathbf{M}}\boldsymbol{\alpha}^{\star}=\lambda_{\max}(\mathbf{P}_{\perp}\bar{\mathbf{M}}\mathbf{P}_{\perp})=\lambda_{\alpha}.
\label{eq:app_lambda_alpha}
\end{equation}
Replacing the random non-centrality by its fractional-delay average gives \eqref{eq:rho_opt} and \eqref{eq:PD_opt}.

\section{Derivation of Lemma~\ref{lem:Pout}}
\label{app:Pout_derivation}

For the desired beam $k_0$, the normalized LoS decision metric is approximated by its mean value
\begin{equation}\small
\gamma_{\mathrm{LoS}}=\frac{N_s^{\mathrm{dwell}}P_{\mathrm{tx}}}{\sigma^2}G_{\mathrm{array}}^{\mathrm{BS}}(\Theta_{\mathrm a})G_{\mathrm{array}}^{\mathrm{BM}}(\Theta_{\mathrm d})\frac{\kappa}{\kappa+1}|\beta_0|^4.
\label{eq:app_gamma_general}
\end{equation}
Using $G_{\mathrm{array}}^{\mathrm{BS}}(\Theta_{\mathrm a})\approx2/\Theta_{\mathrm a}$ and $G_{\mathrm{array}}^{\mathrm{BM}}(\Theta_{\mathrm d})\approx2/\Theta_{\mathrm d}$ gives \eqref{eq:gamma_LoS_explicit}.
For a misaligned beam $k\neq k_0$, let $Z_k$ denote the normalized NLoS-plus-noise decision metric. Its mean is approximated as
\begin{equation}\small
\mathbb{E}[Z_k]=1+\frac{N_s^{\mathrm{dwell}}P_{\mathrm{tx}}}{\sigma^2}G_{\mathrm{array}}^{\mathrm{BS}}(\Theta_{\mathrm a})\bar{G}_{\mathrm{array}}^{\mathrm{BM}}\frac{1}{\kappa+1}|\beta_0|^4.
\label{eq:app_mu_general}
\end{equation}
Using $G_{\mathrm{array}}^{\mathrm{BS}}(\Theta_{\mathrm a})\approx2/\Theta_{\mathrm a}$ and $\bar{G}_{\mathrm{array}}^{\mathrm{BM}}\approx2/\Omega$ gives \eqref{eq:mu_NLoS}.

The effective alignment SINR is defined as
$\eta(\Theta_{\mathrm a};\Theta_{\mathrm d})=\frac{\gamma_{\mathrm{LoS}}(\Theta_{\mathrm a};\Theta_{\mathrm d})}{\bar{\mu}_{\mathrm{NLoS}}(\Theta_{\mathrm a})}$.
Substituting \eqref{eq:gamma_LoS_explicit} and \eqref{eq:mu_NLoS} into it yields \eqref{eq:eaSINR}.

Under the exponential approximation, $Z_k\sim\mathrm{Exp}(\bar{\mu}_{\mathrm{NLoS}})$ for $k\neq k_0$. Therefore,
\begin{equation}\small
\Pr\{Z_k<\gamma_{\mathrm{LoS}}\}=1-\exp\!\left(-\frac{\gamma_{\mathrm{LoS}}}{\bar{\mu}_{\mathrm{NLoS}}}\right)=1-\exp[-\eta(\Theta_{\mathrm a};\Theta_{\mathrm d})].
\label{eq:app_single_bin_success}
\end{equation}
Assuming weak correlation among the $K(\Theta_{\mathrm a})-1$ misaligned beam bins,
\begin{equation}\small
\Pr\{Z_k<\gamma_{\mathrm{LoS}},\forall k\neq k_0\}\approx\left[1-\exp[-\eta(\Theta_{\mathrm a};\Theta_{\mathrm d})]\right]^{K(\Theta_{\mathrm a})-1}.
\label{eq:app_all_bin_success}
\end{equation}
Taking the complement of \eqref{eq:app_all_bin_success} gives \eqref{eq:Pout}.

\section{Proof of Proposition~\ref{prop:Pe2e}}
\label{app:Pe2e_proof}

Let $p_D=P_D(\Theta_{\mathrm d})$. The probability that the first successful discovery occurs at the $n$-th downlink frame is
\begin{equation}\small
\Pr\{N_{\mathrm{disc}}=n\}=(1-p_D)^{n-1}p_D,\quad n=1,2,\ldots.
\label{eq:app_disc_time}
\end{equation}
The trigger is useful if
$T_{\mathrm{coh}}\geq nT_{\mathrm{frame}}+T_{\mathrm{align}}(\Theta_{\mathrm a})$.
Since $T_{\mathrm{coh}}\sim\mathrm{Exp}(\mu)$,
\begin{equation}\small
\Pr\{T_{\mathrm{coh}}\geq nT_{\mathrm{frame}}+T_{\mathrm{align}}(\Theta_{\mathrm a})\}=e^{-\mu[nT_{\mathrm{frame}}+T_{\mathrm{align}}(\Theta_{\mathrm a})]}.
\label{eq:app_coh_tail}
\end{equation}
Hence,
$P_{\mathrm{trig}}=\sum_{n=1}^{\infty}(1-p_D)^{n-1}p_De^{-\mu[nT_{\mathrm{frame}}+T_{\mathrm{align}}(\Theta_{\mathrm a})]}$.

Evaluating the geometric series gives
\begin{equation}\small
P_{\mathrm{trig}}=\frac{p_De^{-\mu[T_{\mathrm{align}}(\Theta_{\mathrm a})+T_{\mathrm{frame}}]}}{1-(1-p_D)e^{-\mu T_{\mathrm{frame}}}}.
\label{eq:app_Ptrig_closed}
\end{equation}
Substituting $p_D=P_D(\Theta_{\mathrm d})$ gives \eqref{eq:Ptrig_closed}. The conditional probability of correct incident-beam selection is
\begin{equation}\small
P_{\mathrm{align}}(\Theta_{\mathrm a};\Theta_{\mathrm d})=1-P_{\mathrm{out}}(\Theta_{\mathrm a};\Theta_{\mathrm d}).
\label{eq:app_Palign}
\end{equation}
Multiplying \eqref{eq:app_Ptrig_closed} and \eqref{eq:app_Palign} gives \eqref{eq:Pe2e_factorized}.

\section{Derivation of the Beamwidth Design Bounds}
\label{app:beamwidth_law_proof}

In the high-discovery-SNR regime, let
$x=\frac{\Theta_{\mathrm d}}{\Omega}$,$A(\Theta_{\mathrm a})=e^{-\mu[T_{\mathrm{align}}(\Theta_{\mathrm a})+T_{\mathrm{frame}}]}$ and $B=e^{-\mu T_{\mathrm{frame}}}$.
Using $P_D(\Theta_{\mathrm d})\approx x$, \eqref{eq:Ptrig_closed} becomes
$P_{\mathrm{trig}}\approx\frac{xA(\Theta_{\mathrm a})}{1-(1-x)B}$.
Imposing $P_{\mathrm{trig}}\geq p_{\mathrm{trig}}$ gives
$xA(\Theta_{\mathrm a})\geq p_{\mathrm{trig}}[1-(1-x)B]$.
Equivalently,
$x[A(\Theta_{\mathrm a})-p_{\mathrm{trig}}B]\geq p_{\mathrm{trig}}(1-B)$.
If $A(\Theta_{\mathrm a})>p_{\mathrm{trig}}B$, then
\begin{equation}\small
x\geq\frac{p_{\mathrm{trig}}(1-B)}{A(\Theta_{\mathrm a})-p_{\mathrm{trig}}B}.
\label{eq:app_x_lower}
\end{equation}
Multiplying \eqref{eq:app_x_lower} by $\Omega$ gives \eqref{eq:theta_d_lower}.

For the alignment constraint, \eqref{eq:Pout} gives
\begin{equation}\small
P_{\mathrm{align}}=\left[1-\exp[-\eta(\Theta_{\mathrm a};\Theta_{\mathrm d})]\right]^{K(\Theta_{\mathrm a})-1}.
\label{eq:app_Palign_exact}
\end{equation}
Using $K(\Theta_{\mathrm a})\approx\Omega/\Theta_{\mathrm a}$, the constraint $P_{\mathrm{align}}\geq p_{\mathrm{align}}$ gives
$1-\exp[-\eta(\Theta_{\mathrm a};\Theta_{\mathrm d})]\geq p_{\mathrm{align}}^{1/(\Omega/\Theta_{\mathrm a}-1)}$.
Thus,
\begin{equation}\small
\eta(\Theta_{\mathrm a};\Theta_{\mathrm d})\geq-\ln\!\left[1-p_{\mathrm{align}}^{1/(\Omega/\Theta_{\mathrm a}-1)}\right].
\label{eq:app_align_ineq_2}
\end{equation}
Substituting \eqref{eq:eaSINR} into \eqref{eq:app_align_ineq_2} yields
\begin{equation}\small
\frac{\kappa|\beta_0|^4/\Theta_{\mathrm d}}{|\beta_0|^4/\Omega+\sigma^2(\kappa+1)\Theta_{\mathrm a}/(4N_s^{\mathrm{dwell}}P_{\mathrm{tx}})}\geq\eta_{\mathrm{req}}(\Theta_{\mathrm a},p_{\mathrm{align}}).
\label{eq:app_align_ineq_3}
\end{equation}
Solving \eqref{eq:app_align_ineq_3} for $\Theta_{\mathrm d}$ gives \eqref{eq:theta_d_upper}. Combining \eqref{eq:theta_d_lower}, \eqref{eq:theta_d_upper}, and the physical interval $\Theta_{\mathrm d}\in[\Theta_{\min}^{\mathrm{BM}},\Omega]$ gives \eqref{eq:theta_d_feasibility}. Since the objective favors smaller $\Theta_{\mathrm a}$, the asymptotic beamwidth rule is \eqref{eq:theta_a_asymptotic}.

%$\hfill\blacksquare$
\ifCLASSOPTIONcaptionsoff
  \newpage
\fi

%\printbibliography %Prints bibliography
\bibliographystyle{IEEEtran}
\bibliography{ref}
% \begin{thebibliography}{32}
% %\bibitem{renzo2019smart} M. D. Renzo, M. Debbah, D.-T. Phan-Huy, A. Zappone, M.-S. Alouini, C. Yuen, et al., ``Smart radio environments empowered by reconfigurable ai meta-surfaces: An idea whose time has come,'' \emph{EURASIP J. Wirel. Commun. Netw.}, vol. 2019, no. 1, pp.~1--20, 2019. 10.1186/s13638-019-1438-9
% \end{thebibliography} 

\end{document}